\documentclass[prd,aps,nofootinbib,onecolumn,floatfix,10 pt]{revtex4}
\pdfoutput=1
\usepackage{amssymb}
\usepackage{amsmath,graphicx,color,epsfig}
\usepackage{slashed}
\begin{document}

\title{Polarized forward-backward asymmetries of lepton pair in $B\rightarrow K_{1}\ell^{+}\ell^{-}$ decay in the presence of New physics}
\author{Faisal Munir$\footnote{faisalmunir@ihep.ac.cn}^{1}$, Saadi Ishaq$\footnote{Saadi.ishaq@uog.edu.pk}^{1,3}$, Ishtiaq Ahmed$\footnote{ishtiaq@ncp.edu.pk}^{2,4}$}
\affiliation{ $^{1}$Center for Future High Energy Physics, Institute of High Energy Physics, Chinese Academy of Sciences, Beijing 100049, China\\
$^{2}$ National Centre for Physics, Quaid-i-Azam University Campus,
Islamabad, 45320 Pakistan\\
$^{3}$Department of Physics, University of Gujrat, Hafiz Hayat
Campus, Gujrat, Pakistan\\
$^{4}$ Laborat\`orio de Física Te\`orica e
Computacional, Universidade Cruzeiro do Sul, 01506-000 S\~ao Paulo,
Brazil}

\date{\today}
\begin{abstract}
Double polarized forward-backward asymmetries in $B\rightarrow
K_{1}(1270,1400)\ell^{+}\ell^{-}$ with $\ell=\mu , \tau$ decays are
studied, using most general non-standard local four-fermi
interactions, where the mass eigenstates $K_{1}(1270)$ and
$K_{1}(1400)$ are the mixture of $^{1}{P}_{1}$ and $^{3}{P}_{1}$
states with the mixing angle $\theta_{K}$. We have calculated the
expressions of nine doubly polarized forward-backward asymmetries
and it is presented that the polarized lepton pair forward-backward
asymmetries are greatly influenced by the new physics. Therefore, these asymmetries are interesting tool to explore the status of new physics in near future, specially at LHC.
\end{abstract}
\pacs{13.20 He, 14.40 Nd}
\maketitle



\section{Introduction}\label{intro}
Rare $B$ decays mediated through the flavor changing neutral current
(FCNC) $b\to s(d)\ell^{+}\ell^{-}$ transitions not only provide a
testing ground for the gauge structure of standard model (SM) but
are also an effective way to look for the physics beyond the SM. As
we know that in SM the Wilson coefficients $C_7, C_9$ and $C_{10}$
of the operators $O_7$, $O_9$ and $O_{10}$ at $\mu=m_b$ are used to
describe the $b\rightarrow s\ell^+\ell^-$ transition. Therefore, in
these transitions, NP effects can be incorporated in two different
ways: one is through new contributions to Wilson coefficients and
the other is via the introduction of new operators in effective
Hamiltonian which are absent in the SM.

Though the decay distribution of inclusive decays such as $B\to
X_{s,d}\ell^{+}\ell^{-}$ is theoretically better understood but
hard to be measured experimentally. In opposite, the exclusive
decays such as $B\to (K,K^{\ast},K_{1},\rho)\ell^{+}\ell^{-}$ are
easy to detect experimentally but are tough to calculate
theoretically as the difficulty lies in describing the hadronic
structure, which are the main source of uncertainties in the
predictions of exclusive rare decays.

 The exploration of physics beyond the SM through various
inclusive $B$ meson decays such as $B\to X_{s,d}\ell^{+}\ell^{-}$
and their corresponding exclusive processes, $B\to
M\ell^{+}\ell^{-}$ with $M = K,K^{\ast},K_{1},\rho$ etc., have
already been studied \cite{1,2,3,bst, 63}. These studies showed that
the above mentioned inclusive and exclusive decays of $B$ meson are
very sensitive to the flavor structure of the SM and provide an
effective way to explore NP effects.

Regarding this precise measurements of different experimental observables for $b\to
s\ell^{+}\ell^{-}$ decay such as branching ratio, forward-backward
asymmetry, various polarization asymmetries of the final state
leptons, etc could be useful in establishing the status of new
physics (NP) in near future, specially at LHC. For this reason, many
exclusive B meson processes based on
$b\rightarrow s\left( d\right) \ell ^{+}\ell ^{-}$ such as  $%
B\rightarrow K\left( K^{\ast }\right) \ell ^{+}\ell ^{-}$\cite{AAli0,Aliev1,Chen,Erkol,WLi,QYan,Kruger1}, $%
B\rightarrow \phi \ell ^{+}\ell ^{-}$\cite{Mohanta}, $B\rightarrow
\gamma \ell ^{+}\ell ^{-}$\cite{Schaudry,Aliev,Yilmaz} and
$B\rightarrow \ell ^{+}\ell ^{-}$\cite{A8}  have already been
studied.

It has been mentioned in \cite{19} that measurement of many
additional observables, would be possible by studying the
simultaneous polarizations of both leptons in the final state, which
in turn would be useful in testing the SM and highlighting new
physics beyond the SM. It should be mentioned here that double
lepton polariztion asymmetries in $B\rightarrow
K^{\ast}\tau^{+}\tau^{-}$\cite{020}, $B\rightarrow
K\ell^{+}\ell^{-}$\cite{021}, $B\rightarrow
\rho\ell^{+}\ell^{-}$\cite{lugu} and $B\rightarrow
K_{1}\ell^{+}\ell^{-}$\cite{V. Bashiry,023} have already been
studied. Along with other observables, forward backward asymmetry is
also an efficient observable to explore NP beyond the SM. In this
regard, double lepton polarization forward-backward asymmertries in
$B\rightarrow K^{\ast}\ell^{+}\ell^{-}$\cite{024,025}, $B\rightarrow
K\ell^{+}\ell^{-}$\cite{gogo}, $B\rightarrow
\rho\ell^{+}\ell^{-}$\cite{tugu} and in
$B_s\rightarrow \gamma \ell^{+}\ell^{-}$\cite{026} have already been
explored. We would like to emphasise here that the situation which
makes $B\rightarrow K_{1}\ell^{+}\ell^{-}$ decay more interesting
than $B\rightarrow K^{\ast}\ell^{+}\ell^{-}$ is the mixing of axial
vector states $K_{1A}$ and $K_{1B}$ which are the $^3P_1$ and
$^1P_1$ states respectively. Therefore, it is also interesting to
see that how polarized forward-backward asymmetries of $B\rightarrow
K_{1}\ell^{+}\ell^{-}$ are influenced in the presence of new
physics. So in the present work polarized forward-backward asymmetry
in the exclusive decay $B\rightarrow K_{1}\ell^{+}\ell^{-}$ are
addressed using most general effective Hamiltonian, including all
forms of possible interactions, similar to the case of $B\rightarrow
K^{\ast}\ell^{+}\ell^{-}$\cite{024} decay.  The physical states
$K_1(1270)$ and $K_1(1400)$ are superposition of the P-wave states
in the following way
\begin{eqnarray}
|K_1(1270)\rangle&=&|K_{1A}\rangle \sin\theta_{K}+|K_{1B}\rangle
\cos\theta_{K}\label{mix1}\\
|K_1(1400)\rangle&=&|K_{1A}\rangle \cos\theta_{K}-|K_{1B}\rangle
\sin\theta_{K}\label{mix2}
\end{eqnarray}
If we define, $y=\sin\theta_{K}$ then above Eqs. become
\begin{eqnarray}
|K_1(1270)\rangle&=&|K_{1A}\rangle y+|K_{1B}\rangle
\sqrt{1-y^2}\notag\\
|K_1(1400)\rangle&=&|K_{1A}\rangle \sqrt{1-y^2}-|K_{1B}\rangle
y\notag
\end{eqnarray}
where the magnitude of the mixing angle $\theta_{K}$ has been
estimated \cite{20} to be $34^\circ\le|\theta_{K}|\le58^\circ$ and the study of
$B\rightarrow K_1(1270)\gamma$ impose the limit \cite{21} on the
mixing angle as
\begin{eqnarray}
\theta_{K}=-(34\pm13)^\circ\label{theta}
\end{eqnarray}
where minus sign of $\theta_K$ is related to the chosen phase of
$K_{1A}$ and $K_{1B}$ \cite{21}.

The manuscript is presented as follows. In sec. \ref{tf}, we devise
our required theoretical framework which is followed by two
Subsections. \ref{mix} and \ref{const}, relating to mixing
 of $K_{1}(1270)$ and $K_{1}(1400)$, form factors and constraints on the coefficients of NP operators used in this study. Sec. \ref{obs}, is devoted
 to analytical calculations and the explicit expressions of doubly polarized forward-backward asymmetries. In Sec. \ref{num},
 we give the numerical analysis with discussion about the observables underconsiderations. We end our work by giving concluding remarks in Sec. \ref{conc}.

\section{Theoretical formalism}\label{tf}

At the quark level $B\rightarrow K_{1}(1270,1400)\ell^{+}\ell^{-}$
decays are induced by the transition $b\rightarrow
s\ell^{+}\ell^{-}$, which in the SM, is described by the following
effective Hamiltonian \cite{22a}
\begin{align}  \label{Amplitude}
&\mathcal{H}_{eff}^{SM}(b \rightarrow s \ell^{+}\ell^{-})
= -\frac{G_{F}\alpha}{\sqrt{2}%
\pi } V_{tb}V_{ts}^{\ast } \bigg\{ C_{9}^{effSM}(\bar{s}\gamma _{\mu
}L b)(\bar{\ell} \gamma ^{\mu
}\ell)\notag\\
&+C_{10}(\bar{s}\gamma _{\mu }L b)(\bar{\ell}\gamma ^{\mu }\gamma
_{5}\ell)
-2m_{b}C_{7}^{effSM}(\bar{s}i\sigma _{\mu \nu }\frac{q^{\nu }}{q^{2}}R b)(\bar{ \ell%
}\gamma ^{\mu }\ell) %
\bigg\}
\end{align}
where $R,L=\left( 1\pm \gamma _{5}\right)/2$ are the projector
operators and $q^2$ is the square of momentum transfer while $C's$
are Wilson coefficients. The effective Wilson coefficient
$C_{9}^{effSM}(\mu )$, can be decomposed into the following three
parts \cite{3, 63}
\begin{equation*}
C_{9}^{effSM}(\mu )=C_{9}(\mu )+Y_{SD}(z,\hat{s})+Y_{LD}(z,\hat{s}),
\end{equation*}
where the parameters $z$ and $\hat{s}$ are defined as $%
z=m_{c}/m_{b},\,\,\,\hat{s}=q^{2}/m_{b}^{2}$. It is important to
mention here that in our numerical calculations of asymmetries and
their average values, we do not include $ Y_{LD}(z,\hat{s})$,
otherwise the asymmetries would be largely effected by the
contributions of $ J/\psi $ and $ \psi (2s)$ resonance around $ s=10
GeV^2 $ and $s=14GeV^2 $ respectively. The explicit expressions for
short-distance contributions $Y_{SD}(z,\hat{s})$ and long distance
contributions $Y_{LD}(z,\hat{s})$ are given in \cite{3, 63}.

New physics effects are explored for $B\rightarrow K_1 l^{+}l^{-}$
channel by considering the most general local four-fermi
interactions. In this regard the total effective Hamiltonian is
given by

\begin{eqnarray}
\mathcal{H}_{eff}=\mathcal{H}^{SM}_{eff}+\mathcal{H}^{VA}_{eff}+\mathcal{H}^{SP}_{eff}+\mathcal{H}^{T}_{eff}\label{13}
\end{eqnarray}
where
\begin{eqnarray}
\mathcal{H}^{VA}_{eff} &=&\frac{G_{F}\alpha }{\sqrt{2}\pi
}V_{ts}^{\ast
}V_{tb}\Big\{ C_{LL}\overline{s}_{L}\gamma ^{\mu }b_{L}\overline{l}%
_{L}\gamma ^{\mu }l_{L}\notag
\\
&+&C_{LR}\overline{s}_{L}\gamma ^{\mu }b_{L}\overline{l}%
_{R}\gamma ^{\mu }l_{R}+C_{RL}\overline{s}_{R}\gamma ^{\mu }b_{R}\overline{l}%
_{L}\gamma ^{\mu }l_{L}\notag
\\
&+&C_{RR}\overline{s}_{R}\gamma ^{\mu }b_{R}\overline{l}%
_{R}\gamma ^{\mu }l_{R}\Big\}  \notag \\
\mathcal{H}^{SP}_{eff} &=&\frac{G_{F}\alpha }{\sqrt{2}\pi
}V_{ts}^{\ast
}V_{tb}\Big\{C_{LRLR}\overline{s}_{L}b_{R}\overline{l}_{L}l_{R}+C_{RLLR}\overline{s%
}_{R}b_{L}\overline{l}_{L}l_{R}\notag
\\
&+&C_{LRRL}\overline{s}_{L}b_{R}\overline{l}%
_{R}l_{L}+C_{RLRL}\overline{s}_{R}b_{L}\overline{l}_{R}l_{L}\Big\}  \notag \\
\mathcal{H}^{T}_{eff} &=&\frac{G_{F}\alpha }{\sqrt{2}\pi
}V_{ts}^{\ast }V_{tb}\Big\{ C_{T}\overline{s}\sigma _{\mu \nu
}b\overline{l}\sigma ^{\mu \nu }l\notag \\
&+&iC_{TE}\epsilon _{\mu \nu \alpha \beta }\overline{l}\sigma ^{\mu
\nu }l\overline{s}\sigma ^{\alpha \beta }b\Big\}
\label{nphysoperator}
\end{eqnarray}%
while $\mathcal{H}^{SM}_{eff}$ is given in Eq. (\ref{Amplitude}) and
$C_X$ are the coefficients of the four-Fermi interactions. Defining
the combinations
\begin{eqnarray}
R_V&=&{1\over2}(C_{LL}+C_{LR}),\quad R_A={1\over2}(C_{LR}+C_{LL})\notag\\
R'_V&=&{1\over2}(C_{RR}+C_{RL}),\quad R'_A={1\over2}(C_{RR}-C_{RL})\notag\\
R_S&=&{1\over2}(C_{LRRL}+C_{LRLR}),\quad R_P={1\over2}(C_{LRLR}-C_{LRRL})\notag\\
R'_S&=&{1\over2}(C_{RLRL}+C_{RLLR}),\quad
R'_P={1\over2}(C_{RLLR}-C_{RLRL})\notag
\end{eqnarray}
where $R_A, R_V, R'_A, R'_V, R_S, R_P, R'_S, R'_P, C_T$ and $C_{TE}$
represents the NP couplings. Using the expression of the effective
Hamiltonian Eq. (\ref{13}) the decay amplitude for $B\rightarrow K_1
l^+ l^-$ is given by

\begin{eqnarray}
&&\mathcal{M}(B\rightarrow K_1 l^+ l^-)=\frac{\alpha
G_F}{2\sqrt2\pi} V_{tb}V^{\ast}_{ts}\notag\\
&&\quad\times\bigg[\langle K_1(p_{K_1},\epsilon)|\bar
s\gamma^{\mu}(1-\gamma_5) b|B(p_B)\rangle\bigg\{(C^{eff}_9+R_V)\bar
l
\gamma_{\mu}l\notag\\
&&+(C_{10}+R_A)\bar l\gamma_{\mu}\gamma_5 l\bigg\}+\langle
K_1(p_{K_1},\epsilon)|\bar s\gamma^{\mu}(1+\gamma_5)
b|B(p_B)\rangle\notag\\
&&\times\bigg\{R'_V
\bar l\gamma_{\mu}l+R'_A\bar l\gamma_{\mu}\gamma_5l\bigg\}\notag\\
&&-2{C^{eff}_7\over s}m_b\langle K_1(p_{K_1},\epsilon)|\bar s
i\sigma_{\mu\nu}q^{\nu}(1+\gamma_5) b|B(p_B)\rangle\bar
l\gamma^{\mu} l\notag
\\
&&+\langle K_1(p_{K_1},\epsilon)|\bar s(1+\gamma_5)
b|B(p_B)\rangle\bigg\{R_S \bar l
l+R_P\bar l\gamma_5 l\bigg\}\notag\\
&&+\langle K_1(p_{K_1},\epsilon)|\bar s\gamma^{\mu}(1-\gamma_5)
b|B(p_B)\rangle\bigg\{R'_S \bar l l+R'_P\bar l\gamma_5
l\bigg\}\notag
\\
&&+2C_T\langle K_1(p_{K_1},\epsilon)|\bar
s\sigma_{\mu\nu} b|B(p_B)\rangle\bar l\sigma^{\mu\nu}l\notag\\
&&+2iC_{TE}\epsilon^{\mu\nu\alpha\beta}\langle
K_1(p_{K_1},\epsilon)|\bar s\sigma_{\mu\nu} b|B(p_B)\rangle\bar
l\sigma_{\alpha\beta}l\bigg]\label{qa}
\end{eqnarray}

Note: One can also consider the new physics contribution coming from
the operator $O'_7=C'_7\bar s\sigma^{\mu\nu}b L F^{\mu\nu}$.
However, in the present study we do not include these effects.

\subsection{Form Factors and Mixing of $K_{1}(1270)-K_{1}(1400)$}\label{mix}

The hadronic matrix elements of quark operators appearing in Eq.
(\ref{qa}) over the meson states, for the exclusive $B\rightarrow
K_{1}(1270,1400)\ell^{+}\ell^{-}$ decays can be parameterized in
terms of the form factors as:

\begin{align}
\left\langle K_{1}(p_{k_1},\varepsilon )\left\vert \bar{s}\gamma
_{\mu }b\right\vert B(p_B)\right\rangle  &=-\Big[\varepsilon _{\mu
}^{\ast }\left(
m_{B}+m_{K_{1}}\right) V_{1}(q^{2})  \notag \\
&-(p_B+p_{k_1})_{\mu }\left( \varepsilon ^{\ast }\cdot q\right) \frac{V_{2}(q^{2})}{%
m_{B}+m_{K_{1}}}\Big]  \notag \\
&+q_{\mu }\left( \varepsilon \cdot q\right)
\frac{2m_{K_{1}}}{q^{2}}\left[ V_{3}(q^{2})-V_{0}(q^{2})\right]\label{tf6} \\
\left\langle K_{1}(p_{k_1},\varepsilon )\left\vert \bar{s}\gamma
_{\mu }\gamma _{5}b\right\vert
B(p_B)\right\rangle  &=\frac{2i\epsilon _{\mu \nu \alpha \beta }}{%
m_{B}+m_{K_{1}}}\varepsilon ^{\ast \nu }p_{k_1}^{\alpha }q^{\beta
}A(q^{2}) \label{tf7}\end{align} where $%
p_B(p_{k_1})$ are the momenta of the $B(K_{1})$ mesons and
$\varepsilon _{\mu }$ correspond to the polarization of the final
state axial vector $K_{1}$ meson. In Eq.(\ref{tf6}) we have
\begin{equation}
V_{3}(q^{2})=\frac{m_{B}+m_{K_{1}}}{2m_{K_{1}}}V_{1}(q^{2})-\frac{m_{B}-m_{K_{1}}}{%
2m_{K_{1}}}V_{2}(q^{2})  \label{tf8}
\end{equation}%
with
\begin{equation*}
V_{3}(0)=V_{0}(0)
\end{equation*}%
Additionally
\begin{eqnarray}
\langle K_{1}\left\vert \bar{s}i\sigma _{\mu \nu }b\right\vert
B\rangle&=&[\varepsilon_\mu ^\ast(p_B+p_{k_1})_{\nu}-\varepsilon_\nu
^\ast(p_B+p_{k_1})_{\mu}]F_{1}(q^{2})\notag\\
&+&\left(\frac{m_B^2-m_{K_1}^2}{q^2}F_2(q^2)\right)(\varepsilon_\mu
^\ast q_\nu-\varepsilon_\nu
^\ast q_\mu)\notag \\
&+&\left(\frac{F_{2}(q^{2})}{q^2}+\frac{F_{3}(q^{2})}{m_{B}^{2}-m_{K_{1}}^{2}}\right)\varepsilon
^{\ast }\cdot
q\notag\\
&\times&\big[(p_B+p_{k_1})_{\nu}q_{\mu}-(p_B+p_{k_1})_{\mu
}q_{\nu}\big]\label{tff9}
\end{eqnarray}
\begin{subequations}
\begin{align}
&\left\langle K_{1}(p_{k_1},\varepsilon )\left\vert \bar{s}i\sigma
_{\mu \nu }q^{\nu }b\right\vert B(p_B)\right\rangle \notag\\
&=\left[ \left( m_{B}^{2}-m_{K_{1}}^{2}\right) \varepsilon _{\mu
}^{\ast}-(\varepsilon^{\ast} \cdot
q)(p_B+p_{k_1})_{\mu }\right] F_{2}(q^{2})  \notag \\
&+(\varepsilon ^{\ast }\cdot q)\left[ q_{\mu }-\frac{q^{2}}{%
m_{B}^{2}-m_{K_{1}}^{2}}(p_B+p_{k_1})_{\mu }\right] F_{3}(q^{2})
\label{tf9}
\end{align}
\begin{equation} \left\langle K_{1}(p_{k_1},\varepsilon )\left\vert \bar{s}i\sigma
_{\mu \nu }q^{\nu }\gamma _{5}b\right\vert
B(p_B)\right\rangle=-2i\epsilon _{\mu \nu \alpha \beta }\varepsilon
^{\ast \nu }p_{K_1}^{\alpha }q^{\beta }F_{1}(q^{2}) \label{tf10}
\end{equation}%
\end{subequations}
with $F_{1}(0)=2F_{2}(0).$ Where Eqs. (\ref{tf9},\ref{tf10}) are
obtained by contracting Eq. (\ref{tff9}) with $q^{\nu}$. Moreover,
the matrix element $\langle K_{1}(p_{k_1},\varepsilon)|\bar{s}(1 \pm
\gamma _{5})b|B(p_B)\rangle$ can be calculated by contracting Eq.
(\ref{tf6}) with $q^{\mu }$ and by making use of the equation of
motions along with Eq. (\ref{tf8}), we have
\begin{eqnarray}
\langle K_{1}(p_{k_1},\varepsilon)|\bar{s}(1 \pm \gamma _{5})b|B(p_B)\rangle = \frac{%
1}{m_{b}}\left\{ \pm 2m_{K_{1}}(\varepsilon ^{*}\cdot
q)V_{0}(q^2)\right\} \notag\\
\label{12}
\end{eqnarray}
where the mass of strange quark has been neglected.

As the physical states $K_{1}(1270)$ and $K_{1}(1400)$ are mixed
states of the $K_{1A}$ and $K_{1B}$ with mixing angle $\theta_{K}$
as defined in Eqs. (\ref{mix1}-\ref{mix2}). The $B\to K_{1}$ form
factors can be parameterized as \cite{V. Bashiry}
\begin{widetext}
\begin{eqnarray}
\left(\begin{array}{c}\langle K_{1}(1270)\vert
\bar{s}\gamma_{\mu}(1-\gamma_{5})b\vert B\rangle\\
\langle K_{1}(1400)\vert \bar{s}\gamma_{\mu}(1-\gamma_{5})b\vert
B\rangle\end{array}\right)&=& M\left(\begin{array}{c}\langle
K_{1A}\vert
\bar{s}\gamma_{\mu}(1-\gamma_{5})b\vert B\rangle\\
\langle K_{1B}\vert \bar{s}\gamma_{\mu}(1-\gamma_{5})b\vert
B\rangle\end{array}\right),\label{m1}\\
\left(\begin{array}{c}\langle K_{1}(1270)\vert
\bar{s}\sigma_{\mu\nu}q^{\mu}(1+\gamma_{5})b\vert B\rangle\\
\langle K_{1}(1400)\vert
\bar{s}\sigma_{\mu\nu}q^{\mu}(1+\gamma_{5})b\vert
B\rangle\end{array}\right)&=& M\left(\begin{array}{c}\langle
K_{1A}\vert
\bar{s}\sigma_{\mu\nu}q^{\mu}(1+\gamma_{5})b\vert B\rangle\\
\langle K_{1B}\vert
\bar{s}\sigma_{\mu\nu}q^{\mu}(1+\gamma_{5})b\vert
B\rangle\end{array}\right),\label{m2}
\end{eqnarray}
\end{widetext}
where the mixing matrix $M$ is
\begin{equation}
M=\left(\begin{array}{cc}\sin\theta_{K}&\cos\theta_{K}\\
\cos\theta_{K}&-\sin\theta_{K}\end{array}\right).\label{m3}
\end{equation}

So the form factors $A^{K_{1}}$, $V_{0,1,2}^{K_{1}}$ and
$F_{0,1,2}^{K_{1}}$ satisfy the following relations
\begin{widetext}
\begin{eqnarray}\left(\begin{array}{c}\frac{A^{K_{1}(1270)}}{m_{B}+m_{K_{1}(1270)}}\\
\frac{A^{K_{1}(1400)}}{m_{B}+m_{K_{1}(1400)}}\end{array}\right) &=&
M\left(\begin{array}{c}\frac{A^{K_{1A}}}{m_{B}+m_{K_{1A}}}\\
\frac{A^{K_{1B}}}{m_{B}+m_{K_{1B}}}\end{array}\right),\label{m4}\\
\left(\begin{array}{c}(m_{B}+m_{K_{1}(1270)})V_{1}^{K_{1}(1270)}\\
(m_{B}+m_{K_{1}(1400)})V_{1}^{K_{1}(1400)}\end{array}\right)&=&
M\left(\begin{array}{c}(m_{B}+m_{K_{1A}})V_{1}^{K_{1A}}\\
(m_{B}+m_{K_{1B}})V_{1}^{K_{1B}}\end{array}\right),\label{m5}\\
\left(\begin{array}{c}\frac{V_{2}^{K_{1}(1270)}}{m_{B}+m_{K_{1}(1270)}}\\
\frac{V_{2}^{K_{1}(1400)}}{m_{B}+m_{K_{1}(1400)}}\end{array}\right)
&=&
M\left(\begin{array}{c}\frac{V_{2}^{K_{1A}}}{m_{B}+m_{K_{1A}}}\\
\frac{V_{2}^{K_{1B}}}{m_{B}+m_{K_{1B}}}\end{array}\right),\label{m6}\\
\left(\begin{array}{c}m_{K_{1}(1270)}V_{0}^{K_{1}(1270)}\\
m_{K_{1}(1400)}V_{0}^{K_{1}(1400)}\end{array}\right) &=&
M\left(\begin{array}{c}m_{K_{1A}}V_{0}^{K_{1A}}\\
m_{K_{1B}}V_{0}^{K_{1B}}\end{array}\right),\label{m7}\\
\left(\begin{array}{c}F_{1}^{K_{1}(1270)}\\
F_{1}^{K_{1}(1400)}\end{array}\right) &=&
M\left(\begin{array}{c}F_{1}^{K_{1A}}\\
F_{1}^{K_{1B}}\end{array}\right),\label{m8}\\
\left(\begin{array}{c}(m^{2}_{B}-m^{2}_{K_{1}(1270)})F_{2}^{K_{1}(1270)}\\
(m^{2}_{B}+m^{2}_{K_{1}(1400)})F_{2}^{K_{1}(1400)}\end{array}\right)
&=&
M\left(\begin{array}{c}(m^{2}_{B}-m^{2}_{K_{1A}})F_{2}^{K_{1A}}\\
(m^{2}_{B}-m^{2}_{K_{1B}})F_{2}^{K_{1B}}\end{array}\right),\label{m9}\\
\left(\begin{array}{c}\frac{F_{3}^{K_{1}(1270)}}{(m^{2}_{B}-m^{2}_{K_{1}(1270)})}\\
\frac{F_{3}^{K_{1}(1400)}}{(m^{2}_{B}-m^{2}_{K_{1}(1400)})}\end{array}\right)
&=&
M\left(\begin{array}{c}\frac{F_{3}^{K_{1A}}}{(m^{2}_{B}-m^{2}_{K_{1A}})}\\
\frac{F_{3}^{K_{1B}}}{(m^{2}_{B}-m^{2}_{K_{1B}})}\end{array}\right),\label{m10}
\end{eqnarray}
\end{widetext}
where we have supposed that $p^{\mu}_{K_{1}(1270),K_{1}(1400)}\simeq
p^{\mu}_{K_{1A},K_{1B}}$.
Using the above matrix elements, the decay
amplitude for $B\rightarrow K_1 l^+ l^-$ can be written as

\begin{eqnarray}
&&\mathcal{M}(B\rightarrow K_1 l^+ l^-)=\frac{\alpha
G_F}{4\sqrt2\pi} V_{tb}V^{\ast}_{ts}(-i)\Bigg[(\bar l
\gamma_{\mu}l)\notag
\\
&&\quad\times\Big\{-2\mathcal{A}\epsilon _{\mu \nu \alpha \beta
}\epsilon ^{\ast \nu }{p}_{K_{1}}^{\alpha }q^{\beta
}-i\mathcal{B}_1\epsilon _{\mu }^{\ast }+i\mathcal{B}_2\varepsilon
^{\ast }\cdot q(p_B+p_{K_1})_\mu\Big\}\notag
\\
&&\quad+(\bar l\gamma_{\mu}\gamma_5 l)\times
\Big\{-2\mathcal{C}_1\epsilon _{\mu \nu \alpha \beta }\epsilon
^{\ast \nu }{p}_{K_{1}}^{\alpha }q^{\beta }-i\mathcal{D}_1\epsilon
_{\mu }^{\ast }\notag\\
&&\quad+i\mathcal{D}_2\varepsilon ^{\ast }\cdot
q(p_B+p_{K_1})_\mu+i\mathcal{D}_0\varepsilon ^{\ast }\cdot q
{q}_{\mu }\Big\}\notag\\
&&\quad+i\mathcal{G}_1(\bar ll)\varepsilon^{\ast }\cdot q+i\mathcal{G}_2(\bar l\gamma_5l)\varepsilon^{\ast }\cdot q\notag\\
&&\quad+4i\mathcal{C}_{TE}\epsilon_{\mu\nu\alpha\beta}(\bar
l\sigma^{\mu\nu}l)\Big\{\mathcal{G}_5\big(\varepsilon^{\ast\alpha
}(p_B+p_{k_1})^{\beta}\notag \\
&&\quad-\varepsilon^{\ast\beta
}(p_B+p_{k_1})^{\alpha}\big)+\mathcal{G}_3(\varepsilon^{\ast\alpha}
q^\beta-\varepsilon^{\ast\beta} q^\alpha)\notag
\\
&&\quad+\mathcal{G}_4\varepsilon ^{\ast }\cdot
q\big[(p_B+p_{k_1})^{\beta}q^{\alpha}-(p_B+p_{k_1})^{\alpha
}q^{\beta}\big] \Big\}\notag
\\
&&\quad+4\mathcal{C}_T(\bar l\sigma_{\mu\nu}l)\Big\{
\mathcal{G}_5\big(\varepsilon^{\ast\mu
}(p_B+p_{k_1})^{\nu}\notag\\
&&\quad-\varepsilon^{\ast\nu
}(p_B+p_{k_1})^{\mu}\big)+\mathcal{G}_3(\varepsilon^{\ast\mu}
q^\nu-\varepsilon^{\ast\nu} q^\mu)\notag \\
&&\quad+\mathcal{G}_4\varepsilon ^{\ast }\cdot
q\big[(p_B+p_{k_1})^{\nu}q^{\mu}-(p_B+p_{k_1})^{\mu
}q^{\nu}\big]\Big\}\Bigg]\label{Amplitude1}
\end{eqnarray}
The auxiliary functions appearing in (\ref{Amplitude1}) can be
written as follows:

\begin{eqnarray*}
\mathcal{A}&=&2(C_9^{eff}+R_V+R_V^{\prime})\frac{A(q^2)}{m_B(1+\hat
k)}+\frac{4m_bC_7^{eff}F_1(q^2)}{q^2}
\\
\mathcal{B}_1&=&2(C_9^{eff}+R_V-R_V^{\prime})m_B(1+\hat
k)V_1(q^2)\notag\\
&&\qquad+4m_bC_7^{eff}(1-\hat k^2)\frac{F_2(q^2)}{({q^2/m_B^2})}
\\
\mathcal{B}_2&=&2(C_9^{eff}+R_V-R_V^{\prime})\frac{V_2(q^2)}{m_B(1+\hat
k)}\notag\\
&&\qquad+\frac{4m_bC_7^{eff}}{q^2}\Bigg[F_2(q^2)+\frac{(q^2/m_B^2)}{(1-\hat
k^2)}F_3(q^2)\Bigg]
\\
\mathcal{C}_1&=&2(C_{10}+R_A+R_A^{\prime})\frac{A(q^2)}{m_B(1+\hat
k)}
\\
\mathcal{D}_1&=&2(C_{10}+R_A-R_A^{\prime})m_B(1+\hat k)V_1(q^2)
\\
\mathcal{D}_2&=&2(C_{10}+R_A-R_A^{\prime})\frac{V_2(q^2)}{m_B(1+\hat
k)}
\\
\mathcal{D}_0&=&\frac{4\hat
k}{m_B}(C_{10}+R_A-R_A^{\prime})\frac{V_3(q^2)-V_0(q^2)}{(q^2/m_B^2)}\notag
\\
\mathcal{G}_1&=&-4(R_S-R_S^{\prime})\frac{\hat
k}{(m_b/m_B)}V_0(q^2)\notag
\\
\mathcal{G}_2&=&-4(R_P-R_P^{\prime})\frac{\hat
k}{(m_b/m_B)}V_0(q^2)\notag
\\
\mathcal{G}_3&=&\frac{m_B^2(1-\hat k^2)}{q^2}F_2(q^2)\notag
\\
\mathcal{G}_4&=&\frac{F_2(q^2)}{q^2}+\frac{F_3(q^2)}{m_B^2(1-\hat
k^2)}\notag
\\
\mathcal{G}_5&=&{F_1(q^2)}
\end{eqnarray*}
where $q=(p_++p_-)=(p_B-p_{K_1})$ and $\hat k\equiv m_{K_1}/m_B$.

\subsection{Phenomenological bounds on NP couplings}\label{const}
In the present paper, we use the constraints on the NP couplings
parameters from A. Kumar {\it{et al}} \cite{DL1}. However, for self
consistency these bounds are given below:

In the absence of $R_{V,A}$ the bounds are

\begin{eqnarray}
|R^{\prime}_V|^2+|R^{\prime}_A|^2\le16.8,
\end{eqnarray}
however these bounds are weakened when we include $R_{V,A}$
\begin{eqnarray}
|R^{\prime}_V|^2+|R^{\prime}_A|^2\le39.7,
\end{eqnarray}

On the other hand the constraints on tensor coupling entirely come
from $B(\bar B\rightarrow X_s\mu^+\mu^-)$ which are
\begin{eqnarray}
|\mathcal{C}_T|^2+4|\mathcal{C}_{TE}|^2\le1.3,
\end{eqnarray}
The limits on scalar and pseudo scalar couplings are extracted from
$B(\bar B^0_s\rightarrow\mu^+\mu^-)$
\begin{eqnarray}
|R_S-R^{\prime}_S|^2+|R_P-R^{\prime}_P|^2\le0.44,
\end{eqnarray}
and from $B(\bar B\rightarrow X_s\mu^+\mu^-)$ \cite{pdg,Aubin} are
\begin{eqnarray}
|R_S|^2+|R_P|^2<45,\qquad R^{\prime}_s=R_S,\qquad R^{\prime}_P=R_P.
\end{eqnarray}

\section{Analytical calculations of doubly polarized forward backward asymmetries}\label{obs}

Now we have all the ingredients to calculate the physical
observables. The double differential decay rate is given
as\cite{DL1}
\begin{eqnarray}
 \frac{d^2\Gamma(B\rightarrow K_1 l^+l^-)}{d\cos\theta ds}=\frac{1}{2m_B}\frac{\rho\sqrt{\lambda}}{(8\pi)^3}|\mathcal{M}|^2
\end{eqnarray}
where $\rho\equiv\sqrt{1-\frac{4m^2}{s}}$ and $\lambda\equiv
m_B^4+m_{k_1}^4+s^2-2m_B^2m_{k_1}^2-2m_B^2s-2m_{k_1}^2s$. By using
the expression of the decay amplitude given in Eq.
(\ref{Amplitude1}) one can get the expression of the dilepton
invariant mass spectrum as

\begin{eqnarray}
  \frac{d\Gamma(B\rightarrow K_1 l^+l^-)}{ds}=\frac{G_F^2\alpha^2m_B}{2^{14}\pi^5}|V_{tb}V_{ts}^\ast|^2\frac{\rho\sqrt{\lambda}}{(8\pi)^3}\delta\label{decay}
\end{eqnarray}
where
\begin{widetext}
\begin{eqnarray*}
\delta&=&4(2m^2+s)\{\frac{8\lambda}{3}|\mathcal{A}|^2+\frac{12m_{K_1}^2s+\lambda}{3m_{K_1}^2s}|\mathcal{B}_1|^2
+\frac{\lambda
t}{3m_{K_1}^2s}\mathcal{R}e(\mathcal{B}_1\mathcal{B}^*_2)
+\frac{\lambda^2}{3m_{K_1}^2s}|\mathcal{B}_2|^2\}
\\
&&+\frac{32\lambda}{3}(s-4m^2)|\mathcal{C}_1|^2+[\frac{4\lambda(2m^2+s)}{3m_{K_1}^2s}+16(s-4m^2)]|\mathcal{D}_1|^2\notag
-8\frac{m^2\lambda}{m_{K_1}^2}\mathcal{R}e(\mathcal{D}_1\mathcal{D}^*_0)-\frac{4\lambda}{3m_{K_1}^2s}[(2m^2+s)
\\
&&\times(m_B^2-m_{K_1}^2)
+s(s-4m^2)]\mathcal{R}e(\mathcal{D}_1\mathcal{D}^*_2)+[6m^2s(2m_B^2+2m_{K_1}^2-s)\notag
+\lambda(2m^2+s)]|\mathcal{D}_2|^2
\\
&&+\frac{8m^2\lambda}{m_{K_1}^2}(m_B^2-m_{K_1}^2)\mathcal{R}e(\mathcal{D}_2\mathcal{D}^*_0)
+\frac{4m^2s\lambda}{m_{K_1}^2}|\mathcal{D}_0|^2
-256m\lambda\mathcal{G}_5\mathcal{R}e(\mathcal{A}\mathcal{C}^*_{TE})+\frac{32m}{m_{K_1}^2}\{\mathcal{G}_4\lambda
t+\mathcal{G}_5(\lambda
\\
&&+12m_{K_1}^2(m_{B}^2-m_{K_1}^2))+\mathcal{G}_3(\lambda+12sm_{K_1}^2)\}\mathcal{R}e(\mathcal{B}_1\mathcal{C}_T^*)
+\frac{32m\lambda}{m_{K_1}^2}(\mathcal{G}_5j+\mathcal{G}_3t+\mathcal{G}_4\lambda)\mathcal{R}e(\mathcal{B}_2\mathcal{C}_T^*)
\\
&&-\frac{4m\lambda}{m_{K_1}^2}\mathcal{R}e(\mathcal{D}_1\mathcal{G}_2^*)
+\frac{4m\lambda(m_{B}^2-m_{K_1}^2)}{m_{K_1}^2}\mathcal{R}e(\mathcal{D}_2\mathcal{G}_2^*)
+\frac{4ms\lambda}{m_{K_1}^2}\mathcal{R}e(\mathcal{D}_0\mathcal{G}_2^*)+\frac{(s-4m^2)\lambda}{m_{K_1}^2}|\mathcal{G}_1|^2
\\
&&+\frac{s\lambda}{m_{K_1}^2}|\mathcal{G}_2|^2
-\frac{256}{3sm_{K_1}^2}\Big[\{12sm_{K_1}^2(8m^2((m_{B}^2+m_{K_1}^2)
-4s)-2\lambda-2s(m_{B}^2+m_{K_1}^2)+s^2)
\\
&&+\lambda(4m^2-s)(s-8{m_{K_1}^2})\}|\mathcal{G}_5|^2
+(4m^2-s)s\{2\lambda
j\mathcal{G}_4\mathcal{G}_5+\lambda^2|\mathcal{G}_4|^2+(24m_{K_1}^2(m_{B}^2-m_{K_1}^2)
\\
&&+13\lambda)\mathcal{G}_3\mathcal{G}_5
+13t\lambda\mathcal{G}_3\mathcal{G}_4+12(\lambda+sm_{K_1}^2)|\mathcal{G}_3|^2\}\Big]|\mathcal{C}_{TE}|^2+\frac{64}{3sm_{K_1}^2}\Big[\{2\mathcal{G}_4\mathcal{G}_5sj
+2\mathcal{G}_3\mathcal{G}_4st
\\
&&+s\lambda|\mathcal{G}_4|^2\}\lambda(8m^2+s)
+\mathcal{G}_3\mathcal{G}_5s\{4m^2(2\lambda+24m_{K_1}^2(m_{B}^2-m_{K_1}^2))+4s(12m_{K_1}^2(m_{B}^2-m_{K_1}^2)+\lambda)\}
\\
&&+|\mathcal{G}_3|^2
4s(3m^2(2\lambda+8sm_{K_1}^2)+s(12sm_{K_1}^2+\lambda))+|\mathcal{G}_5|^2\{4m^2\Big(3s(\lambda-8m_{K_1}^2(s-2(m_{B}^2+m_{K_1}^2)))
\\
&&+(8m_{K_1}^2-s)\lambda\Big)+12sm_{K_1}^2(\lambda+(m_{B}^2-m_{K_1}^2)^2)+\lambda
s(s-8m_{K_1}^2)\}\Big]|\mathcal{C}_T|^2
\end{eqnarray*}
\end{widetext}
where $t\equiv -{m_B}^2+m_{K_1}^2+s$ and $j\equiv -m_{B}^2-3m_{K_1}^2+s$.

we first define the six orthogonal vectors belonging to the
polarizations of $l^-$ and $l^+$ which we denote here by $S_i$ and
$W_i$ respectively where $i=$L, N and T corresponding to the
longitudinally, Normally and transversally polarized lepton $l^\pm$
respectively. \cite{Polarization1,Polarization2,19}

\begin{eqnarray}
 S_L^\mu&\equiv&(0,\textbf{e}_L)=\left(0,\frac{\textbf{p}_-}{|\textbf{p}_-|}\right)\notag\\
 S_N^\mu&\equiv&(0,\textbf{e}_N)=\left(0,\frac{\mathbf{p_{k_1}}\times\textbf{p}_-}{\vert\mathbf{p_{k_1}}\times\textbf{p}_-\vert }\right)
 \notag\\
S_T^\mu&\equiv&(0,\textbf{e}_T)=(0,\textbf{e}_N\times\textbf{e}_L)\label{PV1}
\\
 W_L^\mu&\equiv&(0,\textbf{w}_L)=\left(0,\frac{\textbf{p}_+}{|\textbf{p}_+|}\right)
 \notag\\
 W_N^\mu&\equiv&(0,\textbf{w}_N)=\left(0,\frac{\mathbf{p_{k_1}}\times\textbf{p}_+}{\vert\mathbf{p_{k_1}}\times\mathbf{p}_+ \vert} \right)\notag\\
W_T^\mu&\equiv&(0,\textbf{w}_T)=(0,\textbf{w}_N\times\textbf{w}_L)\label{PV2}
\end{eqnarray}
where $p_+$, $p_-$ and $p_{K_1}$ denote the three momenta vectors of
the final particles $l^+$, $l^-$ and $K_1$ respectively. These
polarization vectors $S_i^\mu(W_i^\mu)$ in Eqs. (\ref{PV1}) and
(\ref{PV2}) are defined in the rest frame of $l^-(l^+)$. When we
apply lorentz boost to bring these polarization vectors from rest
frame of $l^-(l^+)$ to the centre of mass frame of $l^+$ and $l^-$,
only the longitudinal polarization four vector get boosted while the
other two polarization vectors remain unchanged. After this
operation the longitudinal four vector read as

\begin{eqnarray}
 S_L^\mu&=&\left(\frac{|p_-|}{m},\frac{E_l\textbf{p}_-}{m|\textbf{p}_-|}\right)\notag \\
 W_L^\mu&=&\left(\frac{|p_+|}{m},-\frac{E_l\textbf{p}_+}{m|\textbf{p}_+|}\right)\label{PV3}
\end{eqnarray}
To achieve the polarization asymmetries one can use the spin
projector $\frac1 2(1+\gamma_5\slashed S)$ for $l^-$ and for the
$l^+$ spin projector is $\frac1 2(1+\gamma_5\slashed W)$.
Normalized, unpolarized differential forward-backward asymmetry is
defined as

\begin{eqnarray}
\mathcal{A}_{FB}=\frac{\int_{0}^{1}\frac{d^{2}\Gamma }{ds d\cos \theta }%
d\cos \theta -\int_{-1}^{0}\frac{d^{2}\Gamma }{ds d\cos \theta
}d\cos \theta  }{\int_{0}^{1}\frac{d^{2}\Gamma }{ds d\cos \theta }
d\cos \theta+\int_{-1}^{0}\frac{d^{2}\Gamma }{ds d\cos \theta }
d\cos \theta }
\end{eqnarray}
When the spins of both leptons are taken into account, the
$\mathcal{A}_{FB}$ will be a function of the spins of final leptons,
and is defined as

\begin{eqnarray}
\mathcal{A}^{ij}_{FB}=\Big(\frac{d\Gamma}{ds}\Big)^{-1}\Big\{\int_{0}^{1}d\cos
\theta-\int_{-1}^{0}d\cos \theta\Big\}\notag
\\
\times\Big\{\Big[\frac{d^{2}\Gamma(s^{-}=i,s^{+}=j) }{ds d\cos
\theta }-\frac{d^{2}\Gamma(s^{-}=i,s^{+}=-j) }{ds d\cos \theta
}\Big]\notag
\\
-\Big[\frac{d^{2}\Gamma(s^{-}=-i,s^{+}=j) }{ds d\cos
\theta}-\frac{d^{2}\Gamma(s^{-}=-i,s^{+}=-j)}{ds d\cos \theta
}\Big]\Big\}
\end{eqnarray}
\begin{eqnarray}
\mathcal{A}^{ij}_{FB}=\mathcal{A}_{FB}(s^{-}=i,s^{+}=j)-\mathcal{A}_{FB}(s^{-}=i,s^{+}=-j)\notag
\\
-\mathcal{A}_{FB}(s^{-}=-i,s^{+}=j)+\mathcal{A}_{FB}(s^{-}=-i,s^{+}=-j)
\end{eqnarray}
Using these definitions for the double polarized FB asymmetries, we
have found the expressions of numerators as follows:

\begin{eqnarray*}
\mathcal{A}^{LL}_{FB}&=&\sqrt{s\lambda(s-4m^2)}\Big[4\big\{\mathcal{R}e(\mathcal{A}\mathcal{D}^*_1)+\mathcal{R}e(\mathcal{B}_1\mathcal{C}^*_1)\big\}\\
&+&\frac{2m}{m_{K_1}^2s}\big\{t\mathcal{R}e(\mathcal{B}_1\mathcal{G}^*_1)+\lambda\mathcal{R}e(\mathcal{B}_2\mathcal{G}^*_1)\big\}+\frac{64}{s}\\
&\times&\big((m_{B}^2-m_{K_1}^2)\mathcal{G}_5+s\mathcal{G}_3\big)\mathcal{R}e(\mathcal{C}_1
\mathcal{C}_T^*)+\frac{32m}{m_{K_1}^2s}\\
&\times&\Big[(m_{B}^2-5m_{K_1}^2-s)\mathcal{G}_5-\mathcal{G}_3
t-\lambda\mathcal{G}_4\Big]\mathcal{R}e(\mathcal{D}_1\mathcal{C}_{TE}^*)\\
&+&\frac{8m}{m_{K_1}^2}\big(\mathcal{G}_5j+\mathcal{G}_3t+\lambda
\mathcal{G}_4\big)\Big\{\frac{4(m_{B}^2-m_{K_1}^2)}{s}\\
&\times&\mathcal{R}e(\mathcal{D}_2\mathcal{C}^*_{TE})+4\mathcal{R}e(\mathcal{D}_0\mathcal{C}^*_{TE})\\
&+&\frac{1}{m}\Big(\mathcal{R}e(\mathcal{G}_1\mathcal{C}^*_{T})+2\mathcal{R}e(\mathcal{G}_2\mathcal{C}^*_{TE})\Big)\Big\}\Big]
\end{eqnarray*}
\begin{eqnarray*}
\mathcal{A}^{NN}_{FB}&=&\frac{m}{m_{K_1}^2}\sqrt{\frac{\lambda(s-4m^2)}{s}}\Big[-2\big(t\mathcal{R}e(\mathcal{B}_1\mathcal{G}^*_1)
+\lambda\mathcal{R}e(\mathcal{B}_2\mathcal{G}^*_1)\big)\\
&-&32\big(\mathcal{G}_5j+\mathcal{G}_3t+\lambda
\mathcal{G}_4\big)\big[\mathcal{R}e(\mathcal{D}_1\mathcal{C}_{TE}^*)\\
&-&(m_{B}^2-m_{K_1}^2)\mathcal{R}e(\mathcal{D}_2\mathcal{C}^*_{TE})
-s\mathcal{R}e(\mathcal{D}_0\mathcal{C}^*_{TE})\\
&+&\frac{s}{4m}\big\{\mathcal{R}e(\mathcal{G}_1\mathcal{C}^*_{T})
+2\mathcal{R}e(\mathcal{G}_2\mathcal{C}^*_{TE})\big\}\big]\Big]
\end{eqnarray*}
\begin{eqnarray*}
\mathcal{A}^{TT}_{FB}&=&\frac{2m}{m_{K_1}^2}\sqrt{\frac{\lambda(s-4m^2)}{s}}\Big[t\mathcal{R}e(\mathcal{B}_1\mathcal{G}_1^*)
+\lambda\mathcal{R}e(\mathcal{B}_2\mathcal{G}_1^*)\\
&+&16\big(\mathcal{G}_5j+\mathcal{G}_3t+\lambda
\mathcal{G}_4\big)\Big\{\mathcal{R}e(\mathcal{D}_1\mathcal{C}_{TE}^*)\\
&-&(m_{B}^2-m_{K_1}^2)\mathcal{R}e(\mathcal{D}_2\mathcal{C}_{TE}^*)
-s\mathcal{R}e(\mathcal{D}_0\mathcal{C}_{TE}^*)\\
&+&\frac{s}{4m}\big(\mathcal{R}e(\mathcal{G}_1\mathcal{C}_T^*)-2\mathcal{R}e(\mathcal{G}_2\mathcal{C}_{TE}^*)\big)\Big\}\Big]
\end{eqnarray*}
\begin{eqnarray*}
\mathcal{A}^{LT}_{FB}&=&\frac{8\lambda}{3\sqrt{s}}\Big[8(s+4m^2)\mathcal{G}_5\mathcal{R}e(\mathcal{A}\mathcal{C}_{TE}^*)-ms|\mathcal{A}|^2\\
&-&1024m|\mathcal{G}_5|^2|\mathcal{C}_{TE}|^2-4(s-4m^2)\mathcal{G}_5\mathcal{R}e(\mathcal{C}_1\mathcal{C}_{T}^*)\\
&+&\frac{1}{m_{K_1}^2}\Big\{m\Big(|\mathcal{B}_1|^2+t\mathcal{R}e(\mathcal{B}_1\mathcal{B}_2^*)+\lambda|\mathcal{B}_2|^2\Big)\\
&+&(\mathcal{G}_3+\mathcal{G}_5+\mathcal{G}_4t)
\big\{2(s+4m^2)\mathcal{R}e(\mathcal{B}_1\mathcal{C}_{T}^*)\\
&-&4(s-4m^2)\mathcal{R}e(\mathcal{D}_1\mathcal{C}_{TE}^*)\big\}\\
&+&(\mathcal{G}_5j+\mathcal{G}_3t+\mathcal{G}_4\lambda)\big\{2(s+4m^2)\mathcal{R}e(\mathcal{B}_2\mathcal{C}_{T}^*)\\
&-&4(s-4m^2)\mathcal{R}e(\mathcal{D}_2\mathcal{C}_{TE}^*)\big\}\\
&+&64m\big[|\mathcal{G}_5|^2(s-4m_{K_1}^2)+2s(\mathcal{G}_5\mathcal{G}_3
+\mathcal{G}_5\mathcal{G}_4j\\
&+&\mathcal{G}_3\mathcal{G}_4t)+s|\mathcal{G}_3|^2+s\lambda|\mathcal{G}_4|^2\big]|\mathcal{C}_{T}|^2\Big\}\Big]
\end{eqnarray*}
\begin{eqnarray*}
\mathcal{A}^{TL}_{FB}&=&\frac{8\lambda}{3\sqrt{s}}\Big[ms|\mathcal{A}|^2-8(s+4m^2)\mathcal{G}_5\mathcal{R}e(\mathcal{A}\mathcal{C}_{TE}^*)\\
&+&1024m|\mathcal{G}_5|^2|\mathcal{C}_{TE}|^2-4(s-4m^2)\mathcal{G}_5\mathcal{R}e(\mathcal{C}_1\mathcal{C}_{T}^*)\\
&+&\frac{1}{m_{K_1}^2}\Big\{-m\Big(|\mathcal{B}_1|^2+t\mathcal{R}e(\mathcal{B}_1\mathcal{B}_2^*)+\lambda|\mathcal{B}_2|^2\Big)\\
&-&(\mathcal{G}_3+\mathcal{G}_5+\mathcal{G}_4t)
\big\{2(s+4m^2)\mathcal{R}e(\mathcal{B}_1\mathcal{C}_{T}^*)\\
&+&4(s-4m^2)\mathcal{R}e(\mathcal{D}_1\mathcal{C}_{TE}^*)\big\}\\
&-&(\mathcal{G}_5j+\mathcal{G}_3t+\mathcal{G}_4\lambda)\big\{2(s+4m^2)\mathcal{R}e(\mathcal{B}_2\mathcal{C}_{T}^*)\\
&+&4(s-4m^2)\mathcal{R}e(\mathcal{D}_2\mathcal{C}_{TE}^*)\big\}\\
&-&64m\big[|\mathcal{G}_5|^2(s-4m_{K_1}^2)+2s(\mathcal{G}_5\mathcal{G}_3
+\mathcal{G}_5\mathcal{G}_4j\\
&+&\mathcal{G}_3\mathcal{G}_4t)+s|\mathcal{G}_3|^2+s\lambda|\mathcal{G}_4|^2\big]|\mathcal{C}_{T}|^2\Big\}\Big]
\end{eqnarray*}
\begin{eqnarray*}
\mathcal{A}^{NT}_{FB}&=&\frac{m^2\sqrt{\lambda}}{m_{K_1}^2s}\Big[4t\big\{(m_{B}^2-m_{K_1}^2)\mathcal{I}m(\mathcal{B}_1\mathcal{D}^*_2)
-\mathcal{I}m(\mathcal{B}_1\mathcal{D}^*_1)\\
&+&s\mathcal{I}m(\mathcal{B}_1\mathcal{D}^*_0)+\frac{s}{2m}\mathcal{I}m(\mathcal{B}_1\mathcal{G}^*_2)\big\}\\
&+&4\lambda\big\{(m_{B}^2-m_{K_1}^2)\mathcal{I}m(\mathcal{B}_2\mathcal{D}^*_2)-\mathcal{I}m(\mathcal{B}_2\mathcal{D}^*_1)\\
&+&s\mathcal{I}m(\mathcal{B}_2\mathcal{D}^*_0)+\frac{s}{2m}\mathcal{I}m(\mathcal{B}_2\mathcal{G}^*_2)\big\}
\Big]
\end{eqnarray*}
\begin{eqnarray}
\mathcal{A}^{NT}_{FB}=-\mathcal{A}^{TN}_{FB}
\end{eqnarray}
\begin{eqnarray*}
\mathcal{A}^{LN}_{FB}&=&\frac{4\lambda\sqrt{s-4m^2}}{3}\Big[8\mathcal{G}_5\mathcal{I}m(\mathcal{A}\mathcal{C}_T^*)
-m\mathcal{I}m(\mathcal{A}\mathcal{C}^*_1)\\
&+&\frac{1}{m_{K_1}^2s}\Big\{m\big(\mathcal{I}m(\mathcal{B}_1\mathcal{D}^*_1)+t(\mathcal{I}m(\mathcal{B}_1\mathcal{D}^*_2)+\mathcal{I}m(\mathcal{B}_2\mathcal{D}^*_1))\\
&+&\lambda\mathcal{I}m(\mathcal{B}_2\mathcal{D}^*_2)\big)+4s(\mathcal{G}_3+\mathcal{G}_5)\big(2\mathcal{I}m(\mathcal{B}_1\mathcal{C}_{TE}^*)+2t\mathcal{I}m(\mathcal{B}_2\mathcal{C}_{TE}^*))\Big\}\Big]
\end{eqnarray*}
\begin{eqnarray*}
\mathcal{A}^{NL}_{FB}&=&\frac{4\lambda\sqrt{s-4m^2}}{3}\Big[-8\mathcal{G}_5\mathcal{I}m(\mathcal{A}\mathcal{C}_T^*)
-m\mathcal{I}m(\mathcal{A}\mathcal{C}^*_1)\\
&+&\frac{1}{m_{K_1}^2s}\Big\{m\big(\mathcal{I}m(\mathcal{B}_1\mathcal{D}^*_1)+t(\mathcal{I}m(\mathcal{B}_1\mathcal{D}^*_2)+\mathcal{I}m(\mathcal{B}_2\mathcal{D}^*_1))+\lambda\mathcal{I}m(\mathcal{B}_2\mathcal{D}^*_2)\big)\Big\}\\
&-&\frac{8}{m_{K_1}^2}\Big\{(\mathcal{G}_3+\mathcal{G}_5+t\mathcal{G}_4)\mathcal{I}m(\mathcal{B}_1\mathcal{C}_{TE}^*)
+(\mathcal{G}_5j+\mathcal{G}_3t+\mathcal{G}_4\lambda)\mathcal{I}m(\mathcal{B}_2\mathcal{C}_{TE}^*)
\Big\}\Big]
\end{eqnarray*}

Note: It is worthful to mention here we have included short distance
part, $Y_{SD}(z,\hat{s})$, of $C^{effSM}_9$ in our numerical
calculation which contains also the imaginary part, therefore, in
$\mathcal{A}^{NT}_{FB} (\mathcal{A}^{TN}_{FB})$ and $\mathcal{A}^{LN}_{FB} (\mathcal{A}^{NL}_{FB})$ only those terms contribute
which contain auxiliary functions $A, B_1$ and $B_2$.

\section{Numerical Results and Discussion}\label{num}
In this section we examine the effects of different new physics
operators on polarized lepton pair forward-backward asymmetries. For
this purpose, we analyze the behaviour of polarized {\bf\it FB}
asymmetries and their average values in the presence of constraints
on NP couplings that are given in section \ref{const}. Regarding
this, different scenarios for NP Lorentz structure are displayed in
Table IV-VI. Numerical values of different input parameters are
given in Table I, while the SM Wilson coefficients at $\mu=m_b$ are
given in Table II. In addition to calculate the numerical values of
observables under consideration, we have used the light-cone QCD sum
rules form factors \cite{fmf}, summarized in Table \ref{tabel1}. The
momentum dependence dipole parametrization for these form factors
is:
\begin{equation}
\mathcal{T}^{X}_{i}(q^{2})=\frac{\mathcal{T}^{X}_{i}(0)}{1-a_{i}^{X}\left(q^{2}/m^{2}_{B}\right)+b_{i}^{X}\left(q^{2}/m^{2}_{B}\right)^{2}}\label{m11}.
\end{equation}
where $\mathcal{T}$ denotes the $A$, $V$ or $F$ form factors and the
subscript $i$ can take the value 0, 1, 2 or 3. The superscript $X$
belongs to $K_{1A}$ or $K_{1B}$ state.
\begin{table}
\caption{Default values of input parameters used in the calculations
\cite{pdg}}
\begin{tabular}{c}
\hline\hline
$m_{B}=5.28$ GeV, $m_{b}=4.28$ GeV, $m_{\mu}=0.105$ GeV,\\
$m_{\tau}=1.77$ GeV, $f_{B}=0.25$ GeV,
$|V_{tb}V_{ts}^{\ast}|=45\times
10^{-3}$,\\ $\alpha^{-1}=137$, $G_{F}=1.17\times 10^{-5}$ GeV$^{-2}$,\\
$\tau_{B}=1.54\times 10^{-12}$ sec, $m_{K_{1}(1270)}=1.270$ GeV,\\
$m_{K_{1}(1400)}=1.403$ GeV, $m_{K_{1A}}=1.31$ GeV,\\ $m_{K_{1B}}=1.34$ GeV.\\
\hline\hline
\end{tabular}
\end{table}

\begin{table*}[ht]
\centering \caption{The Wilson coefficients $C_{i}^{\mu}$ at the
scale $\mu\sim m_{b}$ in the SM \cite{Ball}.}
\begin{tabular}{cccccccccc}
\hline\hline
$C_{1}$&$C_{2}$&$C_{3}$&$C_{4}$&$C_{5}$&$C_{6}$&$C_{7}$&$C_{9}$&$C_{10}$
\\ \hline
 \ \  1.107 \ \  & \ \  -0.248 \ \  & \ \  -0.011 \ \  & \ \  -0.026 \ \  & \ \  -0.007 \ \  & \ \  -0.031 \ \  & \ \  -0.313 \ \  & \ \  4.344 \ \  & \ \  -4.669 \ \  \\
\hline\hline
\end{tabular}
\label{wc table}
\end{table*}

\begin{table*}[tbp]
\centering \caption{$B\to K_{1A,1B}$ form factors \cite{fmf}, where
$a$ and $b$ are the parameters of the form factors in dipole
parametrization.}
\begin{tabular}{|p{.7in}p{.7in}p{.7in}p{.4in}||p{.7in}p{.7in}p{.7in}p{.4in}|}
\hline \hline
$\mathcal{T}^{X}_{i}(q^{2})$&$\mathcal{T}(0)$&$a$&$b$&$\mathcal{T}^{X}_{i}(q^{2})$&$\mathcal{T}(0)$&$a$&$b$\\
\hline
$V_{1}^{K_{1A}}$&$0.34$&$0.635$&$0.211$&$V_{1}^{K_{1B}}$&$-0.29$&$0.729$&$0.074$\\
$V_{2}^{K_{1A}}$&$0.41$&$1.51$&$1.18$&$V_{2}^{K_{1B}}$&$-0.17$&$0.919$&$0.855$\\
$V_{0}^{K_{1A}}$&$0.22$&$2.40$&$1.78$&$V_{0}^{K_{1B}}$&$-0.45$&$1.34$&$0.690$\\
$A^{K_{1A}}$&$0.45$&$1.60$&$0.974$&$A^{K_{1B}}$&$-0.37$&$1.72$&$0.912$\\
$F_{1}^{K_{1A}}$&$0.31$&$2.01$&$1.50$&$F_{1}^{K_{1B}}$&$-0.25$&$1.59$&$0.790$\\
$F_{2}^{K_{1A}}$&$0.31$&$0.629$&$0.387$&$F_{2}^{K_{1B}}$&$-0.25$&$0.378$&$-0.755$\\
$F_{3}^{K_{1A}}$&$0.28$&$1.36$&$0.720$&$F_{3}^{K_{1B}}$&$-0.11$&$1.61$&$10.2$\\
\hline\hline
\end{tabular}
\label{tabel1}
\end{table*}

Before proceeding to analyze the NP, first we would like to mention
here that the authors of ref \cite{theta,21} concluded that all
observables such as branching ratio, forward backward and single
lepton polarization asymmetries, etc for $B\rightarrow
K_1(1430)\mu^{+}\mu^{-}$ are sensitive to mixing angle $\theta_K$.
In this context, it is interesting to see the dependence of the
values of double lepton polarizations forward-backward asymmetries
on mixing angle $\theta_K$. In this study, we have found that
$\mathcal{A}^{LL}_{FB}$, $\mathcal{A}^{LT}_{FB}$ and
$\mathcal{A}^{TL}_{FB}$ are sensitive to $\theta_K$ for the decay
$B\rightarrow K_1(1430)\mu^{+}\mu^{-}$ as shown in fig 11(a-c) but
not much sensitive for $B\rightarrow K_1(1270)\mu^{+}\mu^{-}$.
Therefore, besides te other observables, the precise measurements of
these asymmetries (for former decay channel) at LHC may also provide
help to put some stringent constraint on the mixing angle $\theta_K$
in near future. However, as it is mentioned in ref. \cite{ishp} that the branching ratio for $K_1(1430)$ is two order suppressed
i.e. $\textit{Br}(B\to K_1(1270)\mu^+\mu^-(\tau^+\tau^-))$ are of
the order of $10^{-6}(10^{-8})$ while $\textit{Br}(B\to
K_1(1430)\mu^+\mu^-(\tau^+\tau^-))$ are of the order of
$10^{-8}(10^{-10})$. For this reason we are not interested in the
results of $B\to K_1(1430)\mu^+\mu^-(\tau^+\tau^-)$.
\begin{table}
\centering \caption{Scenarios for different possible fixed values of
$R_A$, when only $R_A$ and $R_V$ couplings are present}
\begin{tabular}{ccc}
\hline\hline $\text{Scenario}$ & $\hspace{0.5cm}R_A\hspace{0.5cm}$ &
$\hspace{0.5cm}R_V\hspace{0.5cm}$  $\hspace{0.5cm}$\\
\hline
 $\text{S1}$ & $-1.10$ & $-6.5\le R_V\le 1$   \\ \hline
 $\text{S2}$ & $9$ & $-6.5\le R_V\le 1$ \\ \hline\hline
\end{tabular}
\end{table}

\begin{table}
\centering \caption{Scenarios for different possible fixed values of
$R_V'$ and $R_A'$, when only $R_V'$ and $R_A'$ couplings are
present}
\begin{tabular}{ccc}
\hline\hline $\text{Scenario}$ & $\hspace{0.5cm}R_A'\hspace{0.5cm}$
& $\hspace{0.5cm}R_V'\hspace{0.5cm}$ \\ \hline
 $\text{S3}$ & $-3\le R_A'\le 3$ & $0.1$ \\ \hline
 $\text{S4}$ & $-3\le R_A'\le 3$ & $2.75$
 \\ \hline
 $\text{S5}$ & $0.1$ & $-3\le R_V'\le 3$
 \\ \hline
 $\text{S6}$ & $2.75$ & $-3\le R_V'\le 3$  \\ \hline\hline
\end{tabular}
\end{table}

\begin{table}
\centering \caption{Different Scenarios when tensor couplings are
present}
\begin{tabular}{ccc}
\hline\hline $\text{Scenario}$ & $\hspace{0.5cm}C_T\hspace{0.5cm}$ &
$\hspace{0.5cm}C_{TE}\hspace{0.5cm}$\\ \hline
 $\text{S7}$ & $-1.14\le C_T\le 1.14$ & $0$ \\ \hline
 $\text{S8}$ & $0$ & $-0.57\le C_{TE}\le 0.57$ \\ \hline
 $\text{S9}$ & $\pm0.54$ & $-0.5\le C_{TE}\le 0.5$ \\ \hline\hline
\end{tabular}
\end{table}
Now to see the behaviour of double polarized {\bf\it{FB}}
asymmetries under the influence of new physics couplings, we have
drawn the $s$-dependence of these asymmetries in figs. 1-10. In
all these graphs the grey shaded band corresponds to the region of
the SM values of these asymmetries due to uncertainties in mixing
angle $\theta _K$ while dashed line corresponds to the SM value when
the central values of the form factors are taken. In fig. 1 (6) we
present the dependence of $\mathcal{A}^{LL}_{FB}$ on $s$ for the
decay $B\rightarrow K_1(1270)\mu^{+}\mu^{-}$ ($B\rightarrow
K_1(1270)\tau^{+}\tau^{-}$) when only vector type couplings are
switched on. figs. 1a-1c depict the effects of different NP
scenarios presented in tables (IV,V) on $s$ dependence of
$\mathcal{A}^{LL}_{FB}$. These figures show that the zero position
of $\mathcal{A}^{LL}_{FB}$ shifts towards left and right-side of the
corresponding SM value within allowed values of different NP
coefficients. For example fig. 1a depicts scenario S1(see Table IV),
where by fixing the value of $R_A$, three different curves of
$\mathcal{A}^{LL}_{FB}$ are drawn within the allowed range of $R_V$.
It shows that zero position of $\mathcal{A}^{LL}_{FB}$ shifts
towards left and right-side of the corresponding SM value for all
allowed values of $R_V$ in S1 scenario. Similarly figs. 1b and 1c
depict scenarios S4 and S6 given in table V. figs. 2a-2c depict
the effects of tensor interactions (table VI) on $s$ dependence of
$\mathcal{A}^{LL}_{FB}$. For instance fig. 2c shows the case of S9
when both tensor couplings $C_T$ and $C_{TE}$ are present with
opposite polarity. It is important to mention here that only those
scenarios of all NP couplings are shown in figures for which the
zero position of $\mathcal{A}^{LL}_{FB}$ is shifted distinctly in
comparison to that of the zero position in SM. In contrast to
$B\rightarrow K_1(1270)\mu^{+}\mu^{-}$, $\mathcal{A}^{LL}_{FB}$ does
not have zero crossing for $B\rightarrow K_1(1270)\tau^{+}\tau^{-}$.
figs. 6(a-c) depict scenarios S1, S4, S6 and fig. 7(a-c) depict
scenarios S7, S8 and S9 which show, respectively, the possible
effects when only vector and tensor type couplings are present in
$\mathcal{A}^{LL}_{FB}$ for $B\rightarrow
K_1(1270)\tau^{+}\tau^{-}$. In all these scenarios the value of
$\mathcal{A}^{LL}_{FB}$ remains positive in high $s$ region as
predicted by SM value except S7. fig. 7a shows that when tensor
coupling $C_T$ is present only (Scenario S7),
$\mathcal{A}^{LL}_{FB}$ can get the negative values in opposite to
SM prediction. Therefore, if negative values of
$\mathcal{A}^{LL}_{FB}$ are measured in future experiments for
$B\rightarrow K_1(1270)\tau^{+}\tau^{-}$, these results will be
unambiguous indication of existence of new physics beyond the SM
(i.e. existence of tensor type interactions).

In figs. 3(a-d) and 4(a-c), we present the dependence of
$\mathcal{A}^{LT}_{FB}$ on $s$ for muons as final state leptons
while figs. 8(a-c) and 9(a-c) show the dependence of
$\mathcal{A}^{LT}_{FB}$ on $s$ for tauons as final state leptons.
fig 3a (8a) presents S1 (i.e when only $R_A$ and $R_V$ couplings are
present), where three different curves for $\mathcal{A}^{LT}_{FB}$
are plotted by fixing $R_A=-1.10$ and by taking three different
values of $R_V$ within the allowed range (i.e. $-6.5\le R_V\le1$)
for the case of muons (tauons) as final state doubly polarized
leptons. This figure tells us that zero position of
$\mathcal{A}^{LT}_{FB}$ gets shifted towards left and right with
respect to SM zero position for all the different selected values of
$R_V$ with the allowed range while fig. 8a shows the NP effects
when tauons are the final state leptons. One can also see from this
figure that NP effects are significant. Similarly figs. 3b-3d
present the NP effects on $\mathcal{A}^{LT}_{FB}$ when scenarios S2,
S5 and S6 are considered for the decay $B\rightarrow
K_1(1270)\mu^{+}\mu^{-}$. One can also notice from the expressions
given in $\mathcal{A}^{LT}_{FB} (\mathcal{A}^{TL}_{FB})$ that $\mathcal{A}^{LT}_{FB}=-\mathcal{A}^{TL}_{FB}$ when we consider
only vector type couplings. Therefore the effects of vector type
couplings on $\mathcal{A}^{TL}_{FB}$ are same as $\mathcal{A}^{LT}_{FB}$. Moreover, figs. 4a-4c
(9a-9c) present scenarios S7, S8 and S9 for the case of muons
(tauons) as final state leptons when tensor type couplings, $C_T$
and $C_{TE}$, are considered. For instance, fig. 4c represents
scenario S9 (i.e. when both tensor interactions are present) in
which we consider the case when both $C_T$ and $C_{TE}$ are present
with opposite polarity. All these figures show that
$\mathcal{A}^{LT}_{FB}$ is greatly effected by NP couplings in
particular to tensor interactions. Furthermore, for the case of
tauons, when different new physics couplings are switched on, for
some of the cases $\mathcal{A}^{LT}_{FB}$ gets opposite value in
entire high $s$ region as compared to its SM values predictions.

Similar to $\mathcal{A}^{LT}_{FB}$, $s$ dependence of
$\mathcal{A}^{TL}_{FB}$ for different scenarios is shown in figs. 5(a-c) for the decay $B\rightarrow K_1(1270)\mu^{+}\mu^{-}$ while
figs. 10(a-c) present for the case of tauons as final state leptons.
figs.  5(a-c) show the effects of tensor type interactions (S7, S8
and S9). These figures show that all these new physics scenarios
effect the $s$ dependence value of $\mathcal{A}^{TL}_{FB}$
significantly. Additionally, figs. 10(a-c) manifest scenarios S7, S8
and S9 for the case of tauons. Again from these figures we conclude
that different NP couplings modify the value of
$\mathcal{A}^{TL}_{FB}$  significantly in
the high $s$ region.

It is emphasized here that in our analysis only
$\mathcal{A}^{LL}_{FB}$, $\mathcal{A}^{LT}_{FB}$ and
$\mathcal{A}^{TL}_{FB}$ are observed to be considerably effected by
NP couplings of different types. Therefore the other remaining
polarized lepton pair forward-backward asymmetries are not
discussed.

Moreover, we eliminate the dependence of forward-backward polarized
asymmetries on $s$ by performing integration over s and find the
average values of above mentioned asymmetries which are also
experimentally useful tools to explore the new physics. We calculate
the averaged double lepton polarization forward-backward asymmetries
by using the following formula
\begin{eqnarray}
 \langle \mathcal{A}^{ij}_{FB}\rangle=\frac{\int_{4m^2}^{(m_B-m_{K_1})^2}\mathcal{A}^{ij}_{FB}\frac{d\Gamma}{ds}ds}{\int_{4m^2}^{(m_B-m_{K_1})^2}\frac{d\Gamma}{ds}ds}
\end{eqnarray}

As mentioned in Sec. II that in the calculation of average values we
do not include long distance contribution, $Y_{LD}(z,\hat{s})$. Now
we discuss the effects of NP on $\langle
\mathcal{A}^{ij}_{FB}\rangle$, in the following sections.
\subsection{Tensor type interactions present only}
In this section, we discuss the explicit dependence of tensor type couplings on the average values of different polarized forward-backward asymmetries. For this purpose 12e and 12f show the effects of NP tensor and axial tensor operators, respectively, on
$\langle \mathcal{A}^{ij}_{FB}\rangle$ for the case of muons. fig.12e depicts the scenario S7 (see Table-VI. i.e. when $C_T$ present only),
in which $\langle \mathcal{A}^{LL}_{FB}\rangle$ significantly varies
from its SM value. The value of $\langle
\mathcal{A}^{LL}_{FB}\rangle$ increases and reaches to a maximum
value of $\approx 0.21$ and then again decreases within the allowed
range $(-1.14\le C_T\le1.14)$. It is also clear that $\langle
\mathcal{A}^{LL}_{FB}\rangle$ does not change its sign while
$\langle\mathcal{A}^{LT}_{FB}\rangle$ and $\langle
\mathcal{A}^{TL}_{FB}\rangle$ both change their sign in the allowed range. Moreover, $\langle
\mathcal{A}^{LT}_{FB}\rangle$ and
$\langle\mathcal{A}^{TL}_{FB}\rangle$ show opposite trend such that
$\langle \mathcal{A}^{LT}_{FB}\rangle$
($\langle\mathcal{A}^{TL}_{FB}\rangle$) remains positive (negative)
for $(-1.14\le C_T\le-0.05)$ while it becomes negative (positive)
for $(-0.05\le C_T\le1.14)$. All other polarized forward-backward
asymmetries are insignificant for scenario S7. When only second type
of tensor interaction $C_{TE}$ is switched on (Scenario S8), fig.
12f manifest its possible effects on $\langle
\mathcal{A}^{ij}_{FB}\rangle$. It can be easily noted here that only
$\langle\mathcal{A}^{LL}_{FB}\rangle$ does not change its sign while all other change their sign, when
$C_{TE}$ is varied from -0.57 to 0.57. One can also observe that only
$\langle \mathcal{A}^{LL}_{FB}\rangle$, $\langle \mathcal{A}^{TL}_{FB}\rangle$ and $\langle \mathcal{A}^{LT}_{FB}\rangle$ are effected significantly similar to the
case when only $C_T$ type interaction is swithced on.
Similar to figs. 12e and 12f, we plot avaraged double lepton
polarization forward-backward asymmetries in figs. 13e and 13f for
the case of tauons, when only tensor type interactions are present.
fig. 13e shows S7 scenario, where $\langle
\mathcal{A}^{ij}_{FB}\rangle$ is plotted for the allowed range of
$C_T$. From this plot we see that $\langle
\mathcal{A}^{LL}_{FB}\rangle$, $\langle
\mathcal{A}^{LT}_{FB}\rangle$ and $\langle
\mathcal{A}^{TL}_{FB}\rangle$,  are greatly influenced by NP tensor
operator $C_T$ as compared to their SM values, where by signs of
some of these polarized forward-backward asymmetries are flipped as
well. In comparison to this fig. 13f shows even more distinct effects on the
values of all $\langle \mathcal{A}^{ij}_{FB}\rangle$ except $\langle
\mathcal{A}^{TN}_{FB}\rangle$, $\langle
\mathcal{A}^{NT}_{FB}\rangle$, $\langle
\mathcal{A}^{LN}_{FB}\rangle$ and $\langle
\mathcal{A}^{NL}_{FB}\rangle$ (not included), observed for S8,
when only $C_{TE}$ operator is switched on.

\subsection{$R_V$ and $R_A$ couplings present only}
When only $R_V$ and $R_A$ couplings are present, for the case of
muons, figs. 12a and 12b represent scenarios S1 and S2 respectively.
In fig. 12a, When the value of $R_A=-1.10$ is fixed and $R_V$ is
varied in allowed range from -6.5 to 1, $\langle
\mathcal{A}^{LL}_{FB}\rangle$ is drastically changed from its SM
value, while $\langle \mathcal{A}^{LT}_{FB}\rangle$ and $\langle
\mathcal{A}^{TL}_{FB}\rangle$ are also modified appreciably from
their SM values. The value of $\langle \mathcal{A}^{LL}_{FB}\rangle$
remains negative for the values of $R_V$ from -6 to -3 and it
acquires positive values for $(-3\le R_V\le1)$, where the maximum
value $\langle \mathcal{A}^{LL}_{FB}\rangle=0.24$ is observed at
$R_V=1$. It is also clear from this plot that $\langle
\mathcal{A}^{LT}_{FB}\rangle$ and $\langle
\mathcal{A}^{TL}_{FB}\rangle$ follow the opposite pattern, such that
$\langle \mathcal{A}^{LT}_{FB}\rangle$ ($\langle
\mathcal{A}^{TL}_{FB}\rangle$) remains positive (negative) for
$(-6\le R_V\le-1.2)$ and negative (positive) from -1.2 to 1.
Similarly when S2 is considered (fig. 12b), all three double
polarization {\bf\it{FB}} asymmetries, $\langle
\mathcal{A}^{LL}_{FB}\rangle$, $\langle
\mathcal{A}^{LT}_{FB}\rangle$ and $\langle
\mathcal{A}^{TL}_{FB}\rangle$ not only vary in magnitude for the
allowed region of $R_V$ but also change their polarities, where
$\langle \mathcal{A}^{LL}_{FB}\rangle$ becomes positive to negative
at $R_V=-3$ whereas $\langle \mathcal{A}^{LT}_{FB}\rangle$ ($\langle
\mathcal{A}^{TL}_{FB}\rangle$) changes its sign from positive
(negative) to negative (positive) at $R_V\approx-1.2$. All other
averaged polarized {\bf\it{FB}} asymmetries which are left out show
negligible NP effects. When the case of tauons is considered, figs.
13a and 13b, it is observed that presence of couplings $R_V$ and
$R_A$ effect $\langle \mathcal{A}^{LL}_{FB}\rangle$, $\langle
\mathcal{A}^{LT}_{FB}\rangle$ and $\langle\mathcal{A}^{TL}_{FB}\rangle$, significantly. One can observe from
fig. 13a that the magnitude of $\langle
\mathcal{A}^{LL}_{FB}\rangle$, varies significantly within the
allowed range $(-6\le R_V\le1)$ along with the change in polarity of
$\langle \mathcal{A}^{LL}_{FB}\rangle$. While when we consider S2
(fig. 13b), it shows the opposite behaviour for $\langle
\mathcal{A}^{LL}_{FB}\rangle$, while similar behaviour for $\langle
\mathcal{A}^{LT}_{FB}\rangle$ and $\langle
\mathcal{A}^{TL}_{FB}\rangle$ as compared to S1.

\subsection{$R^{\prime}_V$ and $R^{\prime}_A$ couplings present only}
When only $R^{\prime}_V$ and $R^{\prime}_A$ couplings are present,
figs. 12c and 12d depict scenarios s4 and s6 respectively for muons,
while figs. 13c and 13d represent the case of tauons. Again from
these figures, one can observe that only $\langle
\mathcal{A}^{LL}_{FB}\rangle$, $\langle
\mathcal{A}^{LT}_{FB}\rangle$ and $\langle
\mathcal{A}^{TL}_{FB}\rangle$ are considerably effected for the case
of muons as well as for the case of tauons in the presence of $R^{\prime}_V$ and
$R^{\prime}_A$ couplings. In fig. 12c $\langle
\mathcal{A}^{LL}_{FB}\rangle$ and $\langle
\mathcal{A}^{LT}_{FB}\rangle$ acquire only positive sign where by
$\langle \mathcal{A}^{TL}_{FB}\rangle$ acquire only negative sign
for all allowed values of $R^{\prime}_A$. For S6 (fig. 12d),
$\langle \mathcal{A}^{LL}_{FB}\rangle$ increases from 0.03 at
$R^{\prime}_V=-3$ and reaches to a maximum value of $\approx0.23$ at
$R^{\prime}_V=3$ whereas $\langle \mathcal{A}^{LT}_{FB}\rangle$ and
$\langle \mathcal{A}^{TL}_{FB}\rangle$ follow the opposite fashion, compared to s4.
For the case of tauons, fig. 13c show that when we switch on only
$R^{\prime}_V$ and $R^{\prime}_A$, only $\langle
\mathcal{A}^{LL}_{FB}\rangle > {\langle
\mathcal{A}^{LL}_{FB}\rangle}_{SM}$ and $\langle
\mathcal{A}^{LT}_{FB}\rangle > {\langle
\mathcal{A}^{LT}_{FB}\rangle}_{SM}$ is true for all allowed values
of $R^{\prime}_A$, and $\langle \mathcal{A}^{TL}_{FB}\rangle <
{\langle \mathcal{A}^{TL}_{FB}\rangle}_{SM}$ for entire allowed
range $(-3\le R^{\prime}_A \le 3)$ . Similar conclusion can be drawn
from fig. 13d (S6) such as $\langle \mathcal{A}^{LL}_{FB}\rangle >
{\langle \mathcal{A}^{LL}_{FB}\rangle}_{SM}$ for $(0.7\le
R^{\prime}_V \le 3)$ while $\langle \mathcal{A}^{LL}_{FB}\rangle <
{\langle \mathcal{A}^{LL}_{FB}\rangle}_{SM}$ for $(-3\le
R^{\prime}_V \le 0.7)$ . Also polarities of $\langle
\mathcal{A}^{LT}_{FB}\rangle$ and $\langle
\mathcal{A}^{TL}_{FB}\rangle$ are flipped.

\section{Conclusion}\label{conc}
In conclusion, we calculate double polarized {\bf\it{FB}}
asymmetries using most general model independent form of the
effective Hamiltonian including all possible non-standard local
four-fermi interactions. Our analysis shows that similar to the
other observables, polarized {\bf\it{FB}} asymmetries are also
sensitive to the mixing angle $\theta_K$. While considering the
different NP scenarios our analysis exhibit that the averaged double
lepton polarization forward-backward asymmetries are very sensitive
to NP couplings. The key points are as under.

When vector axial-vector couplings are considered for the decay
$B\rightarrow K_1(1270)\mu^{+}\mu^{-}$, only averaged polarized
forward-backward asymmetries, $\langle
\mathcal{A}^{LL}_{FB}\rangle$, $\langle
\mathcal{A}^{LT}_{FB}\rangle$ and $\langle
\mathcal{A}^{TL}_{FB}\rangle$ are effected significantly whereas all
other averaged polarized {\bf\it{FB}} asymmetries are suppressed.
Similarly when only tensor interaction $C_T$ is present again
$\langle \mathcal{A}^{LL}_{FB}\rangle$, $\langle
\mathcal{A}^{LT}_{FB}\rangle$ and $\langle
\mathcal{A}^{TL}_{FB}\rangle$ are modified considerably as compared
to their SM values while $\langle \mathcal{A}^{LL}_{FB}\rangle$,
$\langle \mathcal{A}^{NN}_{FB}\rangle$, $\langle
\mathcal{A}^{TT}_{FB}\rangle$, $\langle
\mathcal{A}^{LT}_{FB}\rangle$ and $\langle
\mathcal{A}^{TL}_{FB}\rangle$ are influenced greatly when only
$C_{TE}$ coupling is present. In similarity to the decay $B\rightarrow
K_1(1270)\mu^{+}\mu^{-}$, when the case of tauons is considered it
is found again that $\langle \mathcal{A}^{LL}_{FB}\rangle$,
$\langle \mathcal{A}^{LT}_{FB}\rangle$ and $\langle
\mathcal{A}^{TL}_{FB}\rangle$ are modified as compared to their
SM values, when either vector axial-vector operators or tensor
interactions of type $C_T$ are present only. Moreover all types of
averaged doubly polarized {\bf\it{FB}} asymmetries except $\langle
\mathcal{A}^{LN}_{FB}, \rangle$, $\langle
\mathcal{A}^{NL}_{FB}\rangle$, $\langle \mathcal{A}^{NT}_{FB}\rangle$ and $\langle \mathcal{A}^{TN}_{FB}\rangle$ are influenced greatly when only
$C_{TE}$ tensor couplings are switched on.

Additionally, the dependence of polarized lepton pair
forward-backward asymmetries $\mathcal{A}^{LL}_{FB}$,
$\mathcal{A}^{LT}_{FB}$ and $\mathcal{A}^{TL}_{FB}$ on $s$ for the
decay $B\rightarrow K_1(1270)\mu^{+}\mu^{-}$ depict the left and
right-side shifting of zero crossing positions of these
forward-backward polarized asymmetries from their corresponding SM
values, when vector axial-vector and tensor type NP operators are
considered. Moreover, signs of some of these polarized {\bf\it{FB}}
asymmetries are also flipped for few allowed values of different NP
couplings. Similar conclusion is drawn for the case of tauons as
final state leptons.

\section*{Acknowledgments}
The authors would like to thank Prof. Fayyazuddin for their valuable
guidance and useful discussions. One of the author I. Ahmed. would like to acknowledge the grant (2013/23177-3) from FAPESP.

\begin{figure*}[ht]
\begin{tabular}{cc}
\hspace{0.6cm}($\mathbf{a}$)&\hspace{1.2cm}($\mathbf{b}$)\\
\includegraphics[scale=0.55]{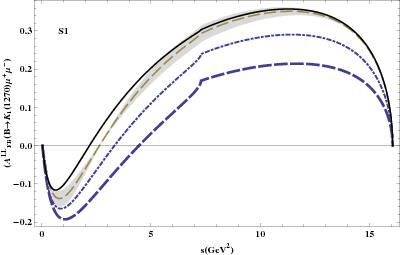} \ \ \
& \ \ \ \includegraphics[scale=0.55]{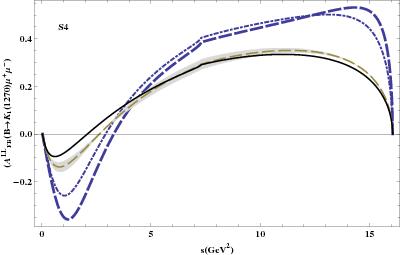}\\
\hspace{0.6cm}($\mathbf{c}$)\\
\includegraphics[scale=0.55]{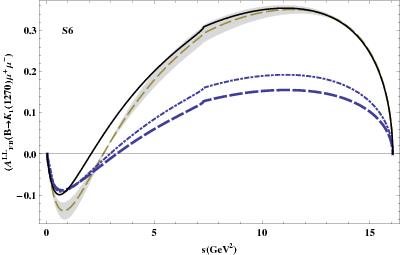} \ \ \
\\
\end{tabular}
\caption{The dependence of $A_{FB}^{LL}$ on $s$ for the decay $B\to
K_1(1270)\mu^+\mu^-$, where the dashed-dotted, solid and dashed
curves in each figure correspond to $A_{FB}^{LL}$ for three
different allowed values of vector type couplings.} \label{lpm}
\begin{tabular}{cc}
\hspace{0.6cm}($\mathbf{a}$)&\hspace{1.2cm}($\mathbf{b}$)\\
\includegraphics[scale=0.55]{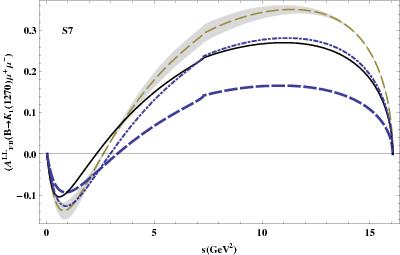} \ \ \
& \ \ \ \includegraphics[scale=0.55]{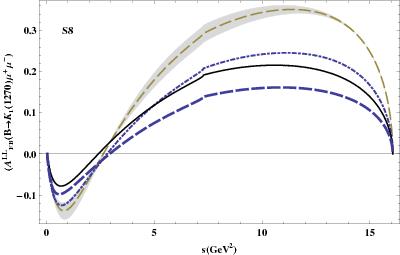}\\
\hspace{0.6cm}($\mathbf{c}$)\\
\includegraphics[scale=0.55]{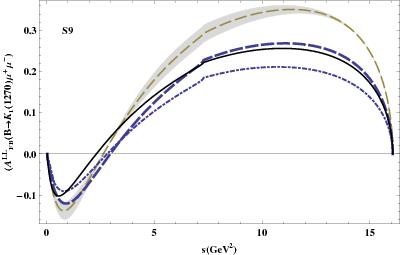} \ \ \
\\
\end{tabular}
\caption{The same as in figure 1, but for tensor type couplings.}
\label{lpm}
\end{figure*}

\begin{figure*}[ht]
\begin{tabular}{cc}
\hspace{0.6cm}($\mathbf{a}$)&\hspace{1.2cm}($\mathbf{b}$)\\
\includegraphics[scale=0.55]{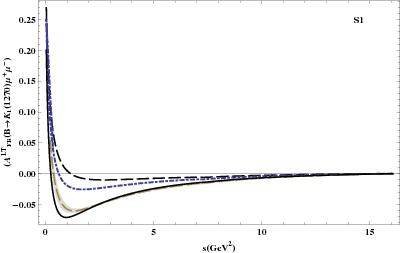} \ \ \
& \ \ \ \includegraphics[scale=0.55]{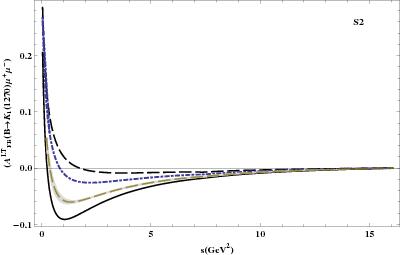}\\
\hspace{0.6cm}($\mathbf{c}$)&\hspace{1.2cm}($\mathbf{d}$)\\
\includegraphics[scale=0.55]{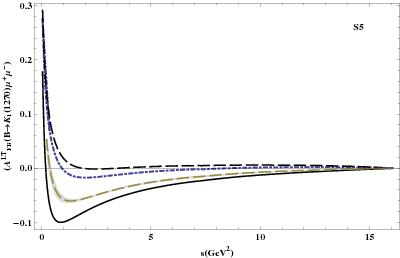} \ \ \
& \ \ \ \includegraphics[scale=0.55]{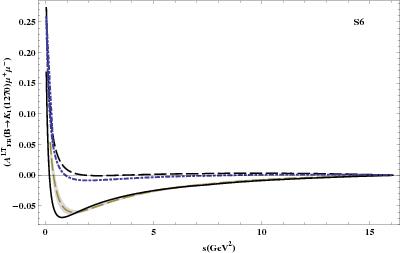}\\
\end{tabular}
\caption{The same as in figure 1, but for $A_{FB}^{LT}$.}
\begin{tabular}{cc}
\hspace{0.6cm}($\mathbf{a}$)&\hspace{1.2cm}($\mathbf{b}$)\\
\includegraphics[scale=0.55]{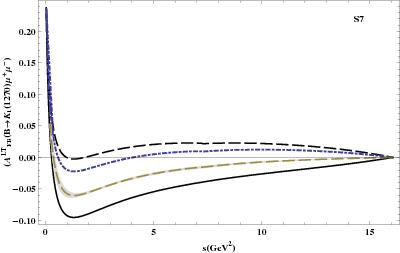} \ \ \
& \ \ \ \includegraphics[scale=0.55]{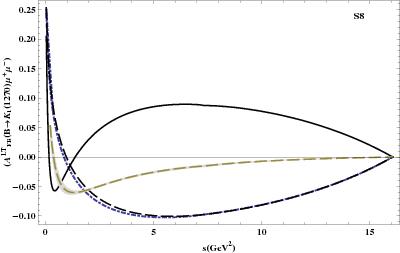}\\
\hspace{0.6cm}($\mathbf{c}$)&\\
\includegraphics[scale=0.55]{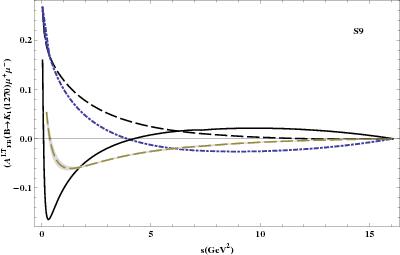} \ \ \
\\
\end{tabular}
\caption{The same as in figure 3, but for tensor type interactions.}
\label{3pm}
\end{figure*}

\begin{figure*}[ht]
\begin{tabular}{cc}
\hspace{0.6cm}($\mathbf{a}$)&\hspace{1.2cm}($\mathbf{b}$)\\
\includegraphics[scale=0.55]{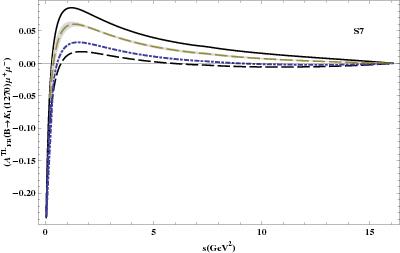} \ \ \
& \ \ \ \includegraphics[scale=0.55]{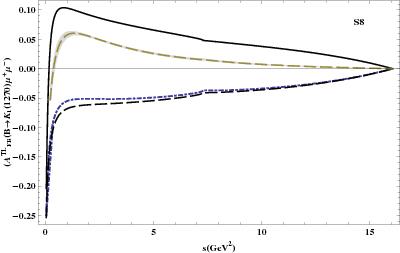}\\
\hspace{0.6cm}($\mathbf{c}$)&\\
\includegraphics[scale=0.55]{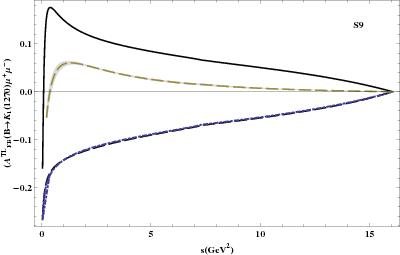} \ \ \
&\\
\end{tabular}
\caption{The dependence of $A_{FB}^{TL}$ on $s$ for the decay $B\to
K_1(1270)\mu^+\mu^-$, where the dashed-dotted, solid and dashed
curves in each figure correspond to $A_{FB}^{TL}$ for three
different allowed values of tensor type couplings.}
\label{lpm}
\end{figure*}

\begin{figure*}[ht]
\begin{tabular}{cc}
\hspace{0.6cm}($\mathbf{a}$)&\hspace{1.2cm}($\mathbf{b}$)\\
\includegraphics[scale=0.55]{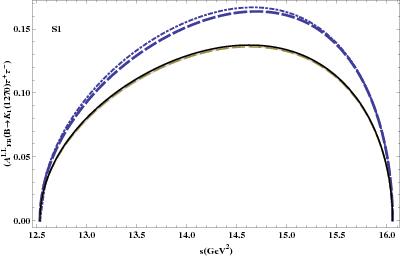} \ \ \
& \ \ \ \includegraphics[scale=0.55]{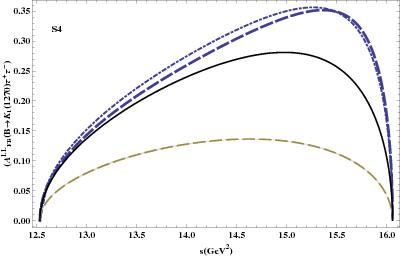}\\
\hspace{0.6cm}($\mathbf{c}$)&\\
\includegraphics[scale=0.55]{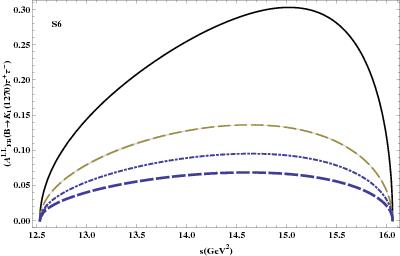} \ \ \
&\\
\end{tabular}
\caption{The same as in figure 1, but for $B\to
K_1(1270)\tau^+\tau^-$.}
\begin{tabular}{cc}
\hspace{0.6cm}($\mathbf{a}$)&\hspace{1.2cm}($\mathbf{b}$)\\
\includegraphics[scale=0.55]{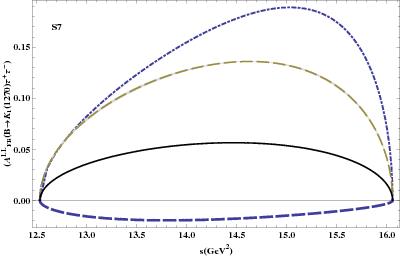} \ \ \
& \ \ \ \includegraphics[scale=0.55]{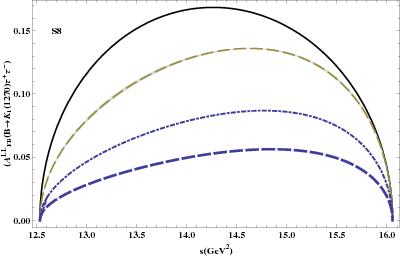}\\
\hspace{0.6cm}($\mathbf{c}$)&\\
\includegraphics[scale=0.55]{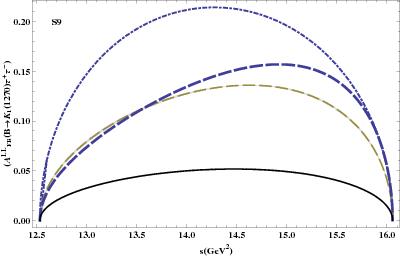} \ \ \
&\\
\end{tabular}
\caption{The same as in figure 2, but for $B\to
K_1(1270)\tau^+\tau^-$.} \label{lpm}
\end{figure*}

\begin{figure*}[ht]
\begin{tabular}{cc}
\hspace{0.6cm}($\mathbf{a}$)&\hspace{1.2cm}($\mathbf{b}$)\\
\includegraphics[scale=0.55]{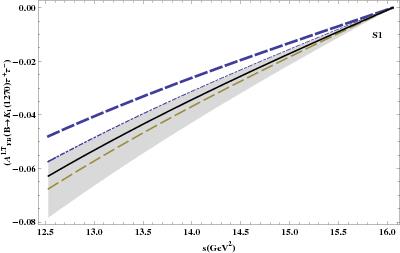} \ \ \
& \ \ \ \includegraphics[scale=0.55]{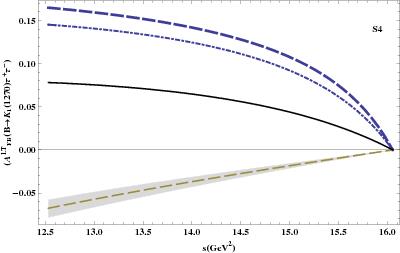}\\
\hspace{0.6cm}($\mathbf{c}$)&\\
\includegraphics[scale=0.55]{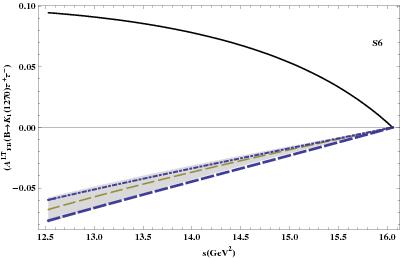} \ \ \
&\\
\end{tabular}
\caption{The same as in figure 3, but for $B\to
K_1(1270)\tau^+\tau^-$.}
\begin{tabular}{cc}
\hspace{0.6cm}($\mathbf{a}$)&\hspace{1.2cm}($\mathbf{b}$)\\
\includegraphics[scale=0.55]{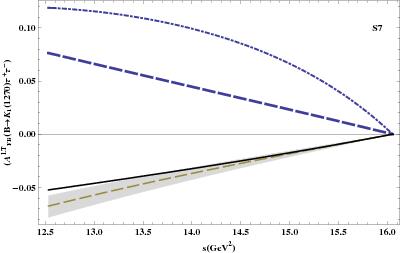} \ \ \
& \ \ \ \includegraphics[scale=0.55]{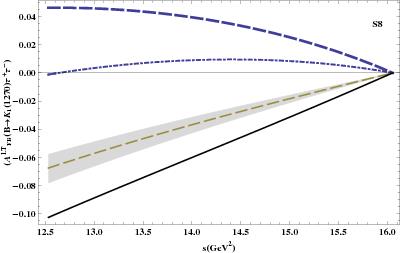}\\
\hspace{0.6cm}($\mathbf{c}$)&\\
\includegraphics[scale=0.55]{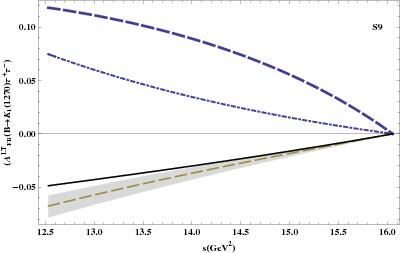} \ \ \
&\\
\end{tabular}
\caption{The same as in figure 4, but for $B\to
K_1(1270)\tau^+\tau^-$.} \label{3pm}
\end{figure*}

\begin{figure*}[ht]
\begin{tabular}{cc}
\hspace{0.6cm}($\mathbf{a}$)&\hspace{1.2cm}($\mathbf{b}$)\\
\includegraphics[scale=0.55]{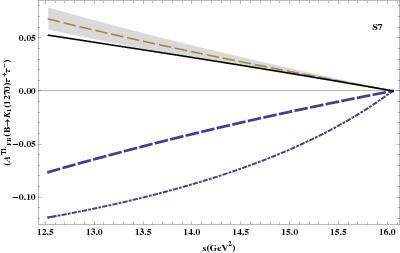} \ \ \
& \ \ \ \includegraphics[scale=0.55]{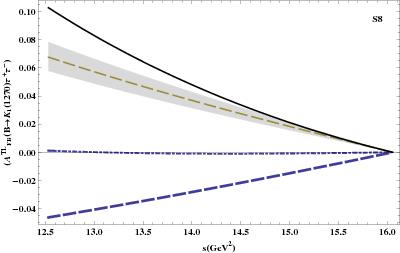}\\
\hspace{0.6cm}($\mathbf{c}$)&\\
\includegraphics[scale=0.55]{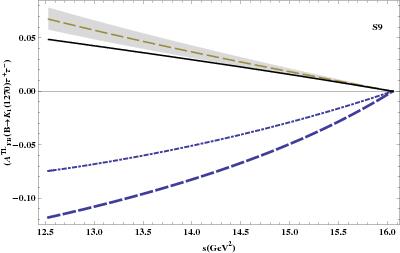} \ \ \
&\\
\end{tabular}
\caption{The same as in figure 5, but for $B\to
K_1(1270)\tau^+\tau^-$.} \label{lpm}
\end{figure*}

\begin{figure*}[ht]
\begin{tabular}{cc}
\hspace{0.6cm}($\mathbf{a}$)&\hspace{1.2cm}($\mathbf{b}$)\\
\includegraphics[scale=0.55]{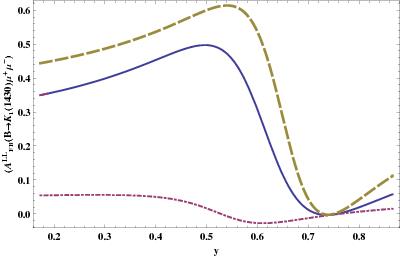} \ \ \
& \ \ \ \includegraphics[scale=0.55]{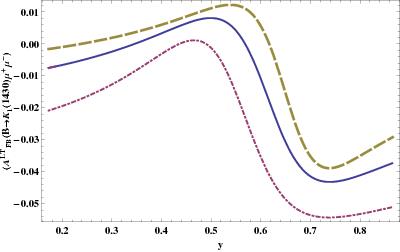}\\
\hspace{0.6cm}($\mathbf{c}$)& \hspace{1.2cm}\\
\includegraphics[scale=0.55]{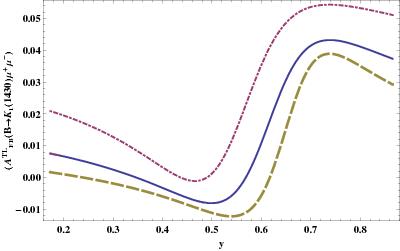} \ \ \
& \ \ \ \\
\end{tabular}
\caption{Dependence of double polarized forward-backward asymmetries
for the decay $B\to K_1(1430)\mu^+\mu^-$ on mixing angle $\theta_K$,
where $y=\sin\theta_K$. The dashed-dotted, solid and dashed angle
dependent curves correspond to $s=3 GeV^2, 5 GeV^2$ and $7 GeV^2$,
respectively.} \label{lpm}
\end{figure*}

\begin{figure*}[ht]
\begin{tabular}{cc}
\hspace{0.6cm}($\mathbf{a}$)&\hspace{1.2cm}($\mathbf{b}$)\\
\includegraphics[scale=0.55]{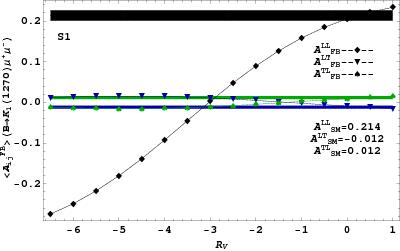} \ \ \
& \ \ \ \includegraphics[scale=0.55]{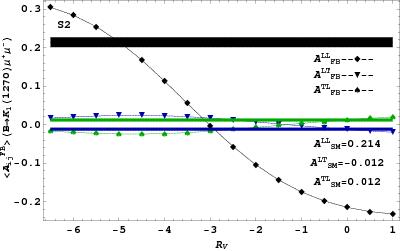}\\
\hspace{0.6cm}($\mathbf{c}$)&\hspace{1.2cm}($\mathbf{d}$)\\
\includegraphics[scale=0.55]{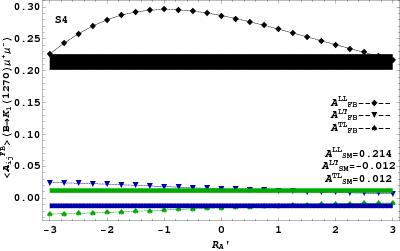} \ \ \
& \ \ \ \includegraphics[scale=0.55]{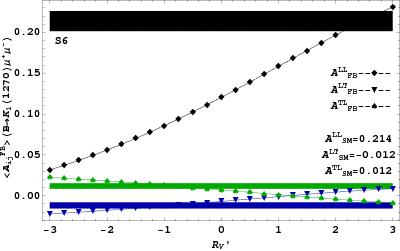}\\
\hspace{0.6cm}($\mathbf{e}$)&\hspace{1.2cm}($\mathbf{f}$)\\
\includegraphics[scale=0.55]{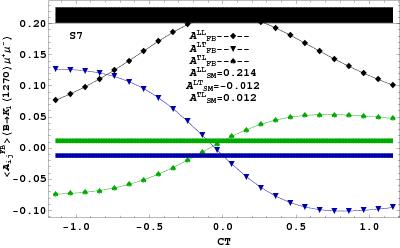} \ \ \
& \ \ \ \includegraphics[scale=0.55]{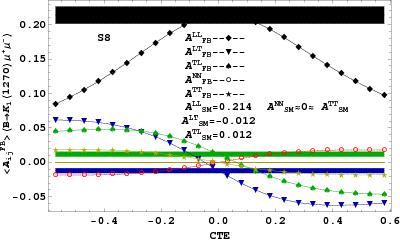}\\
\end{tabular}
\caption{Averaged double lepton Polarized forward backward
asymmetries $\langle\mathcal{A}^{ij}_{FB}\rangle$ for the decay
$B\to K_1(1270)\mu^+\mu^-$ in different scenarios.} \label{lpm}
\end{figure*}

\begin{figure*}[ht]
\begin{tabular}{cc}
\hspace{0.6cm}($\mathbf{a}$)&\hspace{1.2cm}($\mathbf{b}$)\\
\includegraphics[scale=0.55]{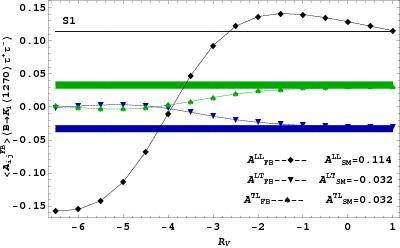} \ \ \
& \ \ \ \includegraphics[scale=0.55]{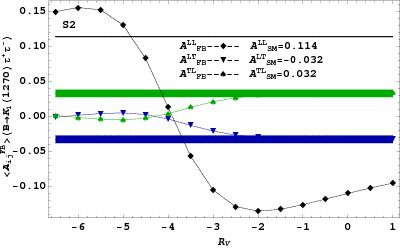}\\
\hspace{0.6cm}($\mathbf{c}$)&\hspace{1.2cm}($\mathbf{d}$)\\
\includegraphics[scale=0.55]{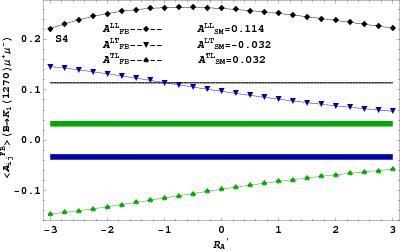} \ \ \
& \ \ \ \includegraphics[scale=0.55]{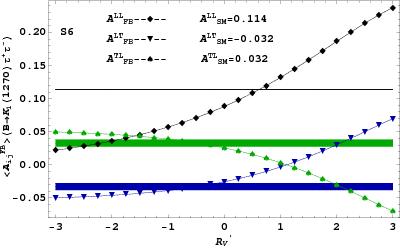}\\
\hspace{0.6cm}($\mathbf{e}$)&\hspace{1.2cm}($\mathbf{f}$)\\
\includegraphics[scale=0.55]{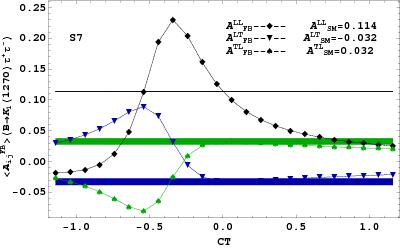} \ \ \
& \ \ \ \includegraphics[scale=0.55]{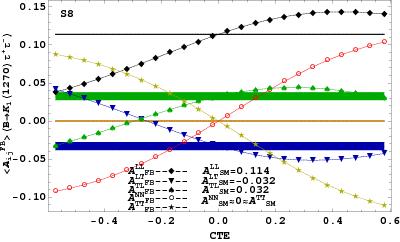}\\
\end{tabular}
\caption{Averaged double lepton Polarized forward backward
asymmetries $\langle\mathcal{A}^{ij}_{FB}\rangle$ for the decay
$B\to K_1(1270)\tau^+\tau^-$ in different scenarios.} \label{lpm}
\end{figure*}


\begin{thebibliography}{99}

\bibitem{1} T. Goto, Y. Okada, Y. Shimizu and M. Tanaha, Phys. Rev. D $\boldsymbol{55}$, 4273
(1997) [arXiv:hep-ph/9609512].

\bibitem{2} S. Bertolini, F. Borzumati, A. Masiero and G. Ridolfi,
Nucl. Phys. B $\boldsymbol{353}$, 591 (1991).

\bibitem{3} C. S. Lim, T. Morozumi and A. I. Sanda, Phys. Lett.
B \textbf{218}, 343 (1989); X.-G. He, T. D. Nguyen and R. R. Volkas, Phys. Rev. D
\textbf{38}, 814 (1988); B. Grinstein, M. J. Savage and M. B. Wise, Nucl. Phys. B \textbf{319} 271 (1989); Y. G. Kim, P. Ko and J. S. Lee, Nucl. Phys.
B \textbf{544}, 64 (1999) [arXiv:hep-ph/9810336]; C.-S. Huang, W.-J. Huo and Y.-L. Wu, Mod.
Phys. Lett. A \textbf{14}, 2453 (1999) [arXiv:hep-ph/9911203].

\bibitem{bst} N. G. Deshpande, J. Trampetic and K. Panose, Phys. Rev.
D \textbf{39}, 1461 (1989) ; P.~J.~O'Donnell and H.~K.~K.~Tung, Phys.
Rev. D \textbf{43}, R2067 (1991); N. Paver and Riazuddin, Phys. Rev.
D \textbf{45}, 978 (1992); W.-S. Hou, R.S.
Willey and A. Soni, Phys. Rev. Lett. \textbf{58}, 1608 (1987) [Erratum-ibid.
\textbf{60}, 2337 (1987)].


\bibitem{63} A. Ali, T. Mannel and T. Morozumi, Phys. Lett. B \textbf{273}, 505 (1991).


\bibitem{AAli0} A. Ali, E. Lunghi, C. Greub and G. Hiller, Phys. Rev. D \textbf{66}, 034002 (2002)
[arXiv:hep-ph/0112300].

\bibitem{Aliev1} T. M. Aliev, M. K. Cakmak and M. Savci, Nucl. Phys. B \textbf{607},
305 (2001) [arXiv:hep-ph/0009133]; T. M. Aliev, A. Ozpineci, M.
Savci and C. Yuce, Phys. Rev. D \textbf{66}, 115006 (2002)
[arXiv:hep-ph/0208128]; T. M. Aliev, A. Ozpineci and M. Savci, Phys.
Lett. B \textbf{511}, 49 (2001) [arXiv:hep-ph/0103261]; T. M. Aliev and
M. Savci, Phys. Lett. B \textbf{481}, 275 (2000)
[arXiv:hep-ph/0003188]; T. M. Aliev, D. A. Demir and M. Savci, Phys. Rev.
D \textbf{62}, 074016 (2000) [arXiv:hep-ph/9912525]; T. M. Aliev,
C. S. Kim and Y. G. Kim, Phys. Rev. D \textbf{62}, 014026 (2000)
[arXiv:hep-ph/9910501]; T. M. Aliev, E.O. Iltan, Phys. Lett. B
\textbf{451}, 175 (1999) [arXiv:hep-ph/9804458].

\bibitem{Chen} C.-H. Chen, C. Q. Geng, Phys. Rev. D \textbf{66}, 034006 (2002)
[arXiv:hep-ph/0207038]; C.-H. Chen, C. Q. Geng, Phys. Rev. D
\textbf{66}, 014007 (2002) [arXiv:hep-ph/0205306].

\bibitem{Erkol} G. Erkol, G. Turan, Nucl. Phys. B
\textbf{635}, 286 (2002) [arXiv:hep-ph/0204219]; E. O. Iltan, G. Turan and I. Turan, J. Phys. G \textbf{28}, 307 (2002)
[arXiv:hep-ph/0106136].

\bibitem{WLi} W.-J. Li, Y.-B. Dai and C.-S. Huang, Eur.
Phys. J. C \textbf{40}, 565 (2005) [arXiv:hep-ph/0410317].



\bibitem{QYan} Q.-S. Yan, C.-S. Huang, W. Liao and S.-H. Zhu, Phys. Rev. D \textbf{62},
094023 (2000) [arXiv:hep-ph/0004262].



\bibitem{Kruger1} F. Kruger, E. Lunghi, Phys. Rev. D \textbf{63}, 014013 (2001)
[arXiv:hep-ph/0008210].

\bibitem{Mohanta} R. Mohanta, A. K. Giri, Phys. Rev. D \textbf{75},
035008 (2007) [arXiv:hep-ph/0611068].

\bibitem{Schaudry} S. R. Choudhury, N. Gaur and N. Mahajan, Phys. Rev. D \textbf{66},
054003 (2002) [arXiv:hep-ph/0203041]; S. R. Choudhury, N. Gaur,
[arXiv:hep-ph/0205076]; S. R. Choudhury, N. Gaur,
[arXiv:hep-ph/0207353].

\bibitem{Aliev} T. M. Aliev, V. Bashiry and M. Savci, Phys. Rev. D \textbf{71}, 035013
(2005) [arXiv:hep-ph/0411327].

\bibitem{Yilmaz} U. O. Yilmaz, B. B. Sirvanli and G. Turan, Nucl. Phys. B \textbf{692}, 249
(2004) [arXiv:hep-ph/0407006]; U. O. Yilmaz, B. B. Sirvanli and G. Turan,
Eur. Phys. J. C \textbf{30}, 197 (2003) [arXiv:hep-ph/0304100].

\bibitem{A8} S. R. Choudhury, N. Gaur, Phys. Lett. B \textbf{451}, 86 (1999)
[arXiv:hep-ph/9810307]; J. K. Mizukoshi, X. Tata and Y. Wang, Phys.
Rev. D \textbf{66}, 115003 (2002) [arXiv:hep-ph/0208078]; A. J. Buras, P. H. Chankowski, J. Rosiek and L. Slawianowska, Nucl. Phys. B \textbf{659}, 3 (2003)
[arXiv:hep-ph/0210145]; A. J. Buras, P. H. Chankowski, J. Rosiek and L. Slawianowska, Phys. Lett. B \textbf{546}, 96 (2002)
[arXiv:hep-ph/0207241].

\bibitem{19} W.~Bensalem, D.~London, N.~Sinha and R.~Sinha, Phys. Rev.  D
{\bf 67}, 034007 (2003) [arXiv:hep-ph/0209228].

\bibitem{020} S. R. Choudhury, N. Gaur, A. S. Cornell and G.C. Joshi, Phys. Rev. D {\bf68}, 054016 (2003) [hep-ph/0304084].
















\bibitem{021} T. M. Aliev, V. Bashiry and M. Savci, Eur. Phys.
J. C \textbf{35}, 197 (2004) [arXiv:hep-ph/0311294].

\bibitem{lugu} T. M. Aliev, V. Bashiry, and M. Savci, Phys. Rev. D \textbf{72},
034031 (2005) [arXiv:hep-ph/0506239].

\bibitem{V. Bashiry} V. Bashiry, JHEP \textbf{0906}, 062 (2009) [arXiv:0902.2578
[hep-ph]].

\bibitem{023} S. Ishaq, F. Munir and I. Ahmed, JHEP $\textbf{07}$, 006 (2013) .


\bibitem{024} T. M. Aliev, V. Bashiry, M. Savci, JHEP \textbf{05}, 037 (2004)  [arXiv:hep-ph/0403282].

\bibitem{025} V. Bashiry, F. Falahati, Phys. Rev. D $\boldsymbol{77}$, 015001 (2008)
[arXiv:0707.3242v1 [hep-ph]].

\bibitem{gogo} S. R. Choudhury, A. S. Cornell, N. Gaur and G. C. Joshi, Phys. Rev. D
\textbf{69}, 054018 (2004) [arXiv:hep-ph/0307276].


\bibitem{tugu} T. M. Aliev, V. Bashiry and M. Savci, Phys. Rev. D
\textbf{73}, 034013 (2006) [arXiv:hep-ph/0504213].





\bibitem{026} T. M. Aliev, V. Bashiry and M. Savci, Eur.
Phys. J. C \textbf{40}, 505 (2005) [arXiv:hep-ph/0412384].


\bibitem{20}  M. Suzuki, Phys. Rev. D $\boldsymbol{47}$, 1252 (1993); L. Burakovsky, J.
T. Goldman, Phys. Rev. D $\boldsymbol{57}$,
2879 (1998) [arXiv:9703271 [hep-ph]]; H. Y. Cheng, Phys. Rev. D
$\boldsymbol{67}$, 094007 (2003) [arXiv: 0301198 [hep-ph]].


\bibitem{21} H. Hatanaka, K. C. Yang, Phys. Rev. D $\boldsymbol{77}$, 094023 (2003) [arXiv:0804.3198 [hep-ph]];
 H. Hatanaka, K. C. Yang, Phys. Rev. D $\boldsymbol{78}$, 074007 (2008) [arXiv:0808.3731 [hep-ph]].


\bibitem{22a} A. J. Buras and M. Munz, Phys,
Rev. D $\boldsymbol{52}$, 186 (1995) [arXiv:hep-ph/9501281]; N. G. Deshpande and J. Trampetic, Phys. Rev. Lett.
$\boldsymbol{60}$, 2583 (1988); M. Misiak, Nucl. Phys. B $\boldsymbol{393}$, 23 (1993) [Erratum-ibid. $\boldsymbol{439}$, 461 (1995)].

\bibitem{DL1} A.~K.~Alok, A.~Dighe, D.~Ghosh, D.~London, J.~Matias, M.~Nagashima and A.~Szynkman,
JHEP {\bf 1002}, 053 (2010) [arXiv:0912.1382 [hep-ph]].

\bibitem{pdg} K.A. Olive \emph{et al.} (Particle Data Group), Chin. Phys. C $\boldsymbol{38}$, 090001 (2014).



\bibitem{Aubin} C. Aubin, [arXiv:0909.2686 [hep-lat]].

\bibitem{Polarization1} F.~Kruger and L.~M.~Sehgal, Phys.
Lett.  B {\bf 380}, 199 (1996) [arXiv:hep-ph/9603237].

\bibitem{Polarization2}  S.~Fukae, C.~S.~Kim and T.~Yoshikawa, Phys. Rev.  D {\bf 61}, 074015 (2000)
[arXiv:hep-ph/9908229].

\bibitem{Ball} A. Ali, P. Ball, L. T. Handoko and G. Hiller, Phys.
Rev. D \textbf{61}, 074024 (2000) [arXiv:hep-ph/9910221].

















\bibitem{fmf} K. C. Yang, Phys. Rev. D \textbf{78}, 034018 (2008) [arXiv:0807.1171].

\bibitem{theta} Y. Li, J. Hua, K.-C. Yang, Eur. Phys. J. C \textbf{71}, 1775 (2011) [arXiv:hep-ph/1107.0630v2].

\bibitem{ishp} A.~Ahmed, I.~Ahmed, M.~Ali Paracha and A.~Rehman,
Phys. Rev. D {\bf 84}, 033010 (2011) [arXiv:1105.3887 [hep-ph]].


\end{thebibliography}
\end{document}